\documentclass[aps,prd,preprint,superscriptaddress,longbibliography,floatfix]{revtex4-1}
\usepackage[T1]{fontenc}
\usepackage[utf8]{inputenc}
\usepackage{lmodern,amsmath,amssymb,bm,booktabs,array,graphicx,microtype,xcolor}
\usepackage[colorlinks=true,citecolor=blue!50!black,linkcolor=blue!50!black,urlcolor=blue!50!black]{hyperref}
\newcommand{\dd}{\mathrm d}
\newcommand{\GeV}{\mathrm{GeV}}
\newcommand{\MeV}{\mathrm{MeV}}
\newcommand{\Ree}{\operatorname{Re}}
\newcommand{\Imm}{\operatorname{Im}}
\newcommand{\Kst}{K^{*0}}
\newcommand{\cL}{\mathcal L}
\newcommand{\figfile}[2]{\IfFileExists{figures/#1}{\includegraphics[width=#2\textwidth,height=0.58\textheight,keepaspectratio]{figures/#1}}{\fbox{Figure #1 is supplied in the source package.}}}
\hypersetup{pdftitle={Magnetic thresholds and spin-resolved spectral moments of phi and Kstar},pdfauthor={Xinyang Wang}}
\begin{document}
\title{Magnetic thresholds and spin-resolved spectral moments of \texorpdfstring{$\phi$ and $K^{*0}$}{phi and Kstar} in a hadronic model}

\author{Lang Yu}
\email{yulang@jlu.edu.cn}
\affiliation{ College of Physics, Jilin University, Changchun 130012, China}

\author{Tao-Tao Sui}
\email{suitaotao@aust.edu.cn}
\affiliation{Center for Fundamental Physics, School of Mechanics and Optoelectronic Physics,
Anhui University of Science and Technology, Huainan, Anhui 232001, China}

\author{Xinyang Wang}
\email{wangxy@aust.edu.cn}
\affiliation{Center for Fundamental Physics, School of Mechanics and Optoelectronic Physics,
Anhui University of Science and Technology, Huainan, Anhui 232001, China}

\begin{abstract}
We study the spin-resolved spectral functions of $\phi$ and $K^{*0}$ in a
thermal hadronic model with a constant magnetic field. We construct the
rest-frame retarded propagators by combining real vacuum matching with
exact Landau levels, connection contacts, and thermal scattering terms.
The resulting stable-daughter calculation is checked against independent
collision kernels and published expressions in common limits. We find
that closed transverse thresholds give rise to narrow peaks with eV-scale
detunings, whose areas must be resolved in the calculation of small spin
spectral moments. We test the leading threshold estimate for the peak area by an
independent integration of the full kernel. Using positive compact
daughter spectra, we then investigate how a finite spectral spread changes
the local threshold structure. The disappearance of zeros of the real part of the inverse propagator
is controlled by the spread relative to the
detuning, whereas the weight integrated over a finite patch can remain
nearly unchanged. We relate this redistribution to smooth accepted moments
through a weighted identity with a bounded Taylor remainder, and evaluate
the complete-window response to three endpoint replacements, including
the patch complement. For the stable-daughter $\phi$ spectra with a fixed smooth weight,
the response at the two smallest sampled fields is negative and compatible with $b^2$
scaling. Hard-window results instead oscillate in sign, and pseudoscalar
mixing substantially changes the smooth response coefficient. These
results quantify the sensitivity to thresholds and spectral projection;
microscopic daughter widths and source-matched experimental spin alignment
require additional input.
\end{abstract}
\maketitle

\section{Introduction}
The polarization of vector mesons provides a way to study the spin
dependence of hadronic interactions. A spectral description relates this
dependence to the retarded self-energy and to the spectral weight of each
spin component. In a magnetic field, the daughter Landau levels introduce
additional thresholds, so the connection between a spectral modification
and a spin observable involves the energy dependence of the propagator
as well as the choice of the measured ensemble. A mass splitting alone
does not determine this connection.

Several approaches have been developed to study vector-meson spectra and
spin alignment in a hadronic medium. Sun, Li and Liu calculated thermal
$\phi$ spectra using chiral and quark--meson interactions and investigated
their relation to gradient-induced alignment \cite{Sun2025}. Hadronic
emission and viscous corrections were considered by Grossi, Palermo and
Zahed \cite{GrossiPalermoZahed2025,GrossiNucleon2025}. Other descriptions
include microscopic spin evolution in a pion gas and anisotropic hadron
coalescence \cite{YinKinetic2024,DongCoalescence2026}. In a complementary
NJL calculation, thermal $\phi$ production and $K^+K^-$ decay were related
to spin alignment through explicit source and decay structures
\cite{ZhuShengHou2025}.

The effects of magnetic fields on hadronic loops have also been studied
in detail. Calculations with charged-pion Landau levels describe neutral
$\rho$ self-energies, decay thresholds, and thermal spectra
\cite{BandyopadhyayMallik2017,RhoThermal2017,RhoGeneral2019}.
Magnetic modifications of the $\phi$ meson have been investigated using
kaon dynamics \cite{Aguirre2019}, while the magnetic response of $K^*$
has been studied on the lattice \cite{KstarLattice2024}. Multipeak
structures associated with thresholds also occur in nonvector spectra
\cite{MeiPion2026}. These studies establish the hadronic and spectral
mechanisms on which the present analysis is based. Our focus is on the
scales governing the narrow threshold structures and on their contribution
to integrated spin spectral moments.

In this paper, we investigate these questions in a specified
$\phi/\Kst$ loop model. We match the real and absorptive parts of the
self-energy on the real energy axis and integrate the threshold
structures without introducing an artificial energy resolution. We then
examine how the spin spectral moments depend on the energy projection and
on analytic matching terms. Comparing the two mesons is useful because
their daughter thresholds differ, and the $K^*$ meson also has
unequal-mass circular channels. As we will show, a pointwise optical check
does not by itself ensure the convergence of an integrated spin ratio.

The threshold analysis identifies two distinct energy scales. Following
thresholds at fixed physical energy, away from a degeneracy with the
parent pole, we obtain $x_*\propto b^2$ for the detuning, whereas the
Landau-level spacing is proportional to $b$. This allows the hierarchy
$x_*\ll h\ll\Delta E_{\rm LL}$: daughter spectral spreads can remove
zeros of the real part of the inverse propagator while adjacent
thresholds remain separated. We study
this regime using a positive two-daughter convolution, with the regular
background fixed by the current-connection model. Explicit weighted
integrals allow us to distinguish a reduction in peak height from a
redistribution of spectral weight. We further connect the local result
to a smooth accepted spin moment by replacing selected analytic
endpoints in the complete inverse propagator and retaining the patch
complement. Separate stable-daughter scans determine the dependence on
the field, acceptance, and matching terms.

The quantities calculated here are defined by the chosen field and
spectral projection. A reconstructed decay angular distribution also
requires production and decay vertices, daughter propagation, and an
observation prescription. This distinction matters when the spectral
ratio differs from $1/3$ by only a few parts in $10^6$.

The paper is organized as follows. Sections~\ref{sec:model} and
\ref{sec:matching} introduce the model and the matched self-energies.
The threshold peaks and their sensitivity to daughter spectral widths
are analyzed in Secs.~\ref{sec:thresholdpeaks} and \ref{sec:width}.
We define the spin spectral moments in Sec.~\ref{sec:moments} and present
the numerical results in Sec.~\ref{sec:numerical}. The implications for
transport are discussed in Sec.~\ref{sec:discussion}, followed by our
conclusions in Sec.~\ref{sec:conclusions}.

\section{Hadronic model and conventions}
\label{sec:model}
We study the vector-meson spectral functions in the rest frame of the
parent. Throughout this paper, we use the metric
$g_{\mu\nu}=\mathrm{diag}(1,-1,-1,-1)$ and natural units. The external
magnetic field is directed along the $z$ axis, and we introduce the
notation $b=eB$ and $\beta=|b|$. The daughter particles have constant
real masses and vanishing chemical potentials. Their thermal occupation
numbers are given by Bose--Einstein distributions at temperature $T$.
The primary decay channels included in the calculation are
$\phi\to K^+K^-,K^0\bar K^0$ and
$\Kst\to K^+\pi^-,K^0\pi^0$.

We fix the normalization of the cubic coupling through the decay
amplitude and the corresponding two-body width,
\begin{equation}
 \mathcal M(V\to P_aP_c)=g\,\epsilon\cdot(p_a-p_c),\qquad
 \Gamma_{ac}(M)=\frac{g^2k^3}{6\pi M^2},\quad
 k=\frac{\sqrt{\lambda(M^2,m_a^2,m_c^2)}}{2M},
 \label{eq:strongwidth}
\end{equation}
where $\lambda(s,a,c)=(s-a-c)^2-4ac$ is the K\"all\'en function.
To complete the interaction, we must also specify the strong four-point
contact, which is not determined by electromagnetic gauging of the cubic
vertex alone. For a charged complex scalar, we obtain the diagonal
$\phi$ interaction from
$|(D_\mu+ig\phi_\mu)P|^2-m_P^2|P|^2$.
For the unequal-mass channel, we introduce the doublet $P=(P_a,P_c)^T$,
whose two components carry the same electric charge, and use the
following scalar Lagrangian:
\begin{equation}
 \cL_P=(\mathcal D_\mu P)^\dagger\mathcal D^\mu P
       -P^\dagger\mathrm{diag}(m_a^2,m_c^2)P,
 \quad
 \mathcal D_\mu=\partial_\mu+ieA_\mu+ig\mathcal V_\mu,
 \quad
 \mathcal V_\mu=\begin{pmatrix}0&V_\mu\\V_\mu^\dagger&0\end{pmatrix}.
 \label{eq:connection}
\end{equation}
Here the conjugate scalar leg represents the physical outgoing
antiparticle. Expanding the covariant derivative gives both the derivative
current vertex and the contact interaction
$g^2V_\mu^\dagger V^\mu(|P_a|^2+|P_c|^2)$.
The neutral $\phi$ channel follows by setting $eA_\mu=0$ in the diagonal
action. In the $K^0\pi^0$ channel, however, the pion is a real field.
We therefore use the corresponding real connection acting on
$(\sqrt2\Ree K^0,\sqrt2\Imm K^0,\pi^0)$, with canonically normalized
real-field kinetic terms. With the phase convention
$i g_n V_\mu\bar K^0\overleftrightarrow\partial^{\mu}\pi^0+\mathrm{H.c.}$,
we obtain the quadratic interaction
\begin{equation}
 \cL_{n,\mathrm{ct}}=g_n^2 V_\mu^\dagger V^\mu
 [\bar K^0K^0+(\pi^0)^2]
 -\frac{g_n^2}{2}\left[V_\mu V^\mu(\bar K^0)^2+
 V_\mu^\dagger V^{\dagger\mu}(K^0)^2\right].
\end{equation}
In the absence of a condensate, the anomalous scalar contractions vanish.
The diagonal $V^\dagger V$ tadpole consequently has the same normalization
in the sum over daughters as in the charged channel. We complete the
vector sector with free Proca terms. Since the mass matrix has unequal
entries, the flavor current is not conserved, and we do not assume an
exact flavor gauge symmetry. Nevertheless, at $\bm p=0$, parity and the
odd dependence on the longitudinal loop momentum imply
$\Sigma^{0z}=0$. Rotational symmetry about the magnetic-field axis also
requires $\Sigma^{0x}=\Sigma^{0y}=0$. The temporal Proca component
therefore decouples from the three spatial modes studied below.

The masses and couplings used in the numerical calculation are listed
in Table~\ref{tab:inputs}. We keep these inputs fixed when defining the
benchmark and do not propagate their experimental uncertainties.

\begin{table}[htbp]\centering\small
\caption{Masses, couplings, and widths used to define the benchmark.
The width inputs follow the PDG listings \cite{PDG2026phi,PDG2026Kstar};
the retained $K^*$ two-body normalization is $47.0\,\MeV$.
The signs of the couplings do not affect the diagonal one-loop spectra.
Masses are given in GeV and radiative couplings in $\GeV^{-1}$.}
\label{tab:inputs}
\begin{tabular}{lrrrrr}\toprule
Parent & $M$ & $g_{\rm charged}$ & $g_{\rm neutral}$ & $\Gamma_{\rm ch}$ (MeV)&$\Gamma_{\rm neu}$ (MeV)\\\midrule
$\phi$ & 1.019460 & 4.516562 & 4.553562 & 2.128749 & 1.427664 \\
$K^{*0}$ & 0.895560 & 4.395891 & -3.108364 & 31.442718 & 15.557282 \\
\midrule
\multicolumn{6}{l}{$m_{K^+}=0.493677$, $m_{K^0}=0.497611$, $m_{\pi^+}=0.13957039$, $m_{\pi^0}=0.1349768$}\\
\multicolumn{6}{l}{$m_\eta=0.547862$, $m_{\eta'}=0.95778$, $\alpha^{-1}=137.035999$}\\
\midrule
\multicolumn{3}{l}{Transition} & \multicolumn{2}{r}{$\Gamma(V\to P\gamma)$ (keV)} & $|h_{VP}|$\\\midrule
\multicolumn{3}{l}{$\phi\to \pi^0\gamma$} & \multicolumn{2}{r}{5.651170} & 0.136006\\
\multicolumn{3}{l}{$\phi\to \eta\gamma$} & \multicolumn{2}{r}{55.491940} & 0.691987\\
\multicolumn{3}{l}{$\phi\to \eta'\gamma$} & \multicolumn{2}{r}{0.264713} & 0.713116\\
\multicolumn{3}{l}{$K^{*0}\to K^0\gamma$} & \multicolumn{2}{r}{116.000000} & 1.268079\\
\bottomrule\end{tabular}\end{table}

In the primary $\phi$ calculation, we omit the three-pion channel and do
not represent its width by an additional constant imaginary term. To
examine the effect of known pseudoscalar transitions, we consider them
separately. The interactions for real and complex fields are written as
\begin{align}
 \cL_{\phi P\gamma}&=e\sum_{P=\pi^0,\eta,\eta'}h_{\phi P}
 \widetilde F_{\mu\nu}(\partial^\mu P)\phi^\nu,\\
 \cL_{K^*K\gamma}&=e h_{K^*K}\widetilde F_{\mu\nu}
 (\partial^\mu\bar K^0)V^\nu+\mathrm{H.c.}
\end{align}
With these conventions, the radiative width and the mixing contribution
to the longitudinal self-energy are given by
\begin{equation}
 \Gamma(V\to P\gamma)=\frac{\alpha h_{VP}^2k_\gamma^3}{3},\qquad
 \Sigma^{VP}_0(s)=b^2\sum_P\frac{h_{VP}^2s}{s-m_P^2}.
 \label{eq:vpmix}
\end{equation}
At zero spatial momentum, the mixing affects only the longitudinal
mode. The radiative couplings listed in the input table fix the matching
of the leading operators. They do not, however, determine all possible
off-shell electromagnetic terms.

We set the coefficients of additional local electromagnetic operators to
zero in the primary loop benchmark. To illustrate the dependence on one
such contribution, we also vary
\begin{equation}
 \delta\Sigma^{\rm loc}_0=2M\xi b^2,\qquad
 \delta\Sigma^{\rm loc}_\pm=0,\qquad [\xi]=\GeV^{-3}.
 \label{eq:xi}
\end{equation}
This variation selects a particular combination of the common and tensor
terms quadratic in the magnetic field in the chosen basis; it does not
represent an uncertainty distribution. An independent Pauli term is also
allowed for the complex neutral $K^*$, but its coefficient is held at
zero. Local terms with derivatives and additional medium operators are
beyond the truncation considered here.

For each spatial mode $m=0,+,-$, we define the inverse retarded
propagator, absorptive self-energy, and spectral function by
\begin{equation}
 \begin{aligned}
 F_m(E)&=E^2-M_0^2-\Sigma_m^R(E),\qquad Q_m(E)=-\Imm\Sigma_m^R(E),\\
 A_m(E)&=-2\Imm F_m^{-1}(E)=\frac{2Q_m}{(\Ree F_m)^2+Q_m^2}.
 \end{aligned}
 \label{eq:spectral}
\end{equation}
We choose the magnetic-field direction as the spin quantization axis.
Reversing the sign of $b$ interchanges the two circular channels.
For $K^{*0}$, the particle labels must also be distinguished from the
physical circular labels of the antiparticle.

\section{Vacuum and thermal self-energies}
\label{sec:matching}
\subsection{Vacuum subtractions and magnetic matching}
We first fix the vacuum subtraction scheme at zero magnetic field.
The inverse propagator from the two-body channels is subtracted at the
real reference mass according to
\begin{equation}
 F_{\rm vac}^{(0)}(s)=s-M^2-
 \left[\Sigma_{\rm vac}^{(0)}(s)-\Ree\Sigma_{\rm vac}^{(0)}(M^2)
 -(s-M^2)\Ree\Sigma_{\rm vac}^{(0)\prime}(M^2)\right].
 \label{eq:onshell}
\end{equation}
Since the subtraction coefficients are real, the physical cuts of the
retained channels remain in the inverse propagator. The reference mass
specifies the subtraction scheme and is not obtained from a separate fit
to a complex pole.

To express the magnetic contribution in dispersive form, we introduce
$\widehat Q^{\rm vac}_{\beta,m}(t)=Q^{\rm vac}_{\beta,m}(\sqrt t)$,
where the argument is the squared energy. The charged vacuum remainder
$R_{\beta,m}$, with two subtractions at $s_*<0$, is then defined by
\begin{equation}
 R_{\beta,m}(s)=-\frac{(s-s_*)^2}{\pi}
 \int_{s_{\rm th}}^\infty\dd t\,
 \frac{\widehat Q^{\rm vac}_{\beta,m}(t)}{(t-s_*)^2(t-s-i0)}.
 \label{eq:remainder}
\end{equation}
The dispersive remainder leaves the subtraction value and slope
undetermined. We calculate both quantities from the same connection
model that defines the interaction. Introducing the shorthand notation
$a=m_a^2$, $c=m_c^2$, $u=x(1-x)$,
$\mathcal D=xa+(1-x)c-s_*u$ and $h(y)=y/\sinh y-1$, we obtain
\begin{align}
 \Delta\Sigma_L^{\rm vac}(s_*,\beta)
 &=-\frac{g^2}{8\pi^2}\int_0^1\dd x\int_0^\infty
 \frac{\dd\tau}{\tau^2}e^{-\mathcal D\tau}h(\beta\tau)
 +\Delta C_{\rm vac},\\
 \Delta C_{\rm vac}
 &=\frac{g^2}{16\pi^2}\int_0^\infty\frac{\dd\tau}{\tau^2}
 (e^{-a\tau}+e^{-c\tau})h(\beta\tau).
 \label{eq:common}
\end{align}
Differentiation with respect to $s_*$ inserts a factor of $u\tau$ into
the bubble term alone. The common contact is independent of the external
energy, and its magnetic difference is finite. As a check, expanding
$h$ at small field gives
\begin{equation}
 \lim_{\beta\to0}\frac{\Delta\Sigma_L^{\rm vac}}{\beta^2}
 =\frac{g^2}{48\pi^2}\int_0^1\frac{\dd x}{\mathcal D}
 -\frac{g^2}{96\pi^2}\left(\frac1a+\frac1c\right).
 \label{eq:weakcheck}
\end{equation}
At the matching point, we separate the tensor and odd contributions at
finite field by defining
$H_\beta=(\Sigma_L-\Sigma_{T\rm av})/\beta^2$ and
$A_\beta=(\Sigma_+-\Sigma_-)/(2\beta)$.
The circular components therefore satisfy
$\Delta\Sigma_\pm=\Delta\Sigma_L-\beta^2H_\beta\pm\beta A_\beta$,
with an identical relation for their slopes. The heat kernels needed to
calculate the charged bubble are
\begin{equation}
 K_L(y)=\frac{y}{\sinh y},\qquad
 K_\pm(y,x)=\frac{y^2e^{\mp(1-2x)y}}{\sinh^2 y}.
\end{equation}
Using $f=g^2/(8\pi^2)$ and $d_x=1-2x$, we find the following finite
matching differences:
\begin{align}
 H_\beta(s_*)&=-\frac f{\beta^2}\int_0^1\dd x\int_0^\infty
 \frac{\dd\tau}{\tau^2}e^{-\mathcal D\tau}
 \left[\frac y{\sinh y}-\frac{y^2\cosh(d_xy)}{\sinh^2 y}\right],\\
 A_\beta(s_*)&=-f\int_0^1\dd x\,d_x\log\frac{\mathcal D}{\mu_r^2}
 +\frac f\beta\int_0^1\dd x\int_0^\infty\frac{\dd\tau}{\tau^2}
 e^{-\mathcal D\tau}
 \left[\frac{y^2\sinh(d_xy)}{\sinh^2y}-d_xy\right],
 \qquad y=\beta\tau .
\end{align}
The result is independent of the scale $\mu_r$ because
$\int_0^1d_x\dd x=0$. The slopes follow by differentiating at $s_*$.
Combining these matching terms with the dispersive remainder gives the
complete magnetic shift,
\begin{equation}
 \Delta\Sigma_m^{\rm vac}(s)=R_{\beta,m}(s)-R_0(s)
 +\Delta\Sigma_m^{\rm vac}(s_*)
 +(s-s_*)\Delta\Sigma_m^{{\rm vac}\prime}(s_*).
 \label{eq:fullmatching}
\end{equation}
Here the subtraction of $R_0$ removes the charged zero-field loop that
is already included in Eq.~\eqref{eq:onshell}.

For numerical evaluation, we express the vacuum Landau sums in terms of
gamma and digamma functions. On the real energy rim, their recurrence
relations separate a finite number of logarithmic integrals and rational
principal values. The special functions in the remaining smooth
integrand have positive arguments. The relevant recurrences include
\begin{align}
 \log|\Gamma(z)|&=\log\Gamma(z+N)-\sum_{j=0}^{N-1}\log|z+j|,\\
 q\psi(1+q)&=q\psi(1+q+N)-N+\sum_{j=1}^{N}\frac{j}{q+j}.
\end{align}
This representation requires neither an imaginary energy resolution
nor a cutoff on the vacuum Landau sum. At a branch point, we evaluate
the appropriate one-sided limit instead of the function at the endpoint.

\subsection{Thermal cuts and collision rates}
We next include the thermal contributions from pair production and
scattering. For $b>0$, the longitudinal Landau pairs are characterized by
$A_n=a+(2n+1)\beta$, $C_n=c+(2n+1)\beta$ and $n\geq0$.
For the circular modes, the corresponding quantities are
$A_{rm}=a+(2r-m)\beta$, $C_{rm}=c+(2r+m)\beta$, with $r\geq1$ and
$m=\pm1$. In the charged $K^{*0}$ channel, we use $P_a=K^+$ and
$P_c=\pi^+$. Then $m=+1$ corresponds to $(n_a,n_c)=(r-1,r)$, whereas
$m=-1$ corresponds to $(r,r-1)$. The outgoing negative pion is
represented by the conjugate leg. Interchanging these circular labels
does not affect the tensor ratios.

For a given Landau pair, we introduce the energies
$E_a=\sqrt{k_z^2+A}$ and $E_c=\sqrt{k_z^2+C}$, together with
$S=E_a+E_c$, $d=E_a-E_c$ and $n_i=n_B(E_i)$.
After removing the vacuum part of the numerator, the thermal bubble
correction takes the form
\begin{equation}
 \delta\Sigma_{m,\rm bub}^T(z)=
 -\frac{g^2\beta}{\pi^2}\sum_\ell\int_0^\infty\dd k_z\,
 \frac{v_m}{E_aE_c}
 \left[\frac{S(n_a+n_c)}{S^2-z^2}
       +\frac{d(n_c-n_a)}{d^2-z^2}\right],
 \quad v_0=k_z^2,\quad v_\pm=\beta r.
 \label{eq:thermal}
\end{equation}
The same connection model gives the thermal contact term
\begin{equation}
 C_T(\beta)=\frac{g^2\beta}{2\pi^2}\sum_{n=0}^\infty
 \int_0^\infty\dd k_z
 \left[\frac{n_B(\sqrt{k_z^2+a+(2n+1)\beta})}
 {\sqrt{k_z^2+a+(2n+1)\beta}}+(a\to c)\right].
\end{equation}
The second term in Eq.~\eqref{eq:thermal} describes thermal scattering.
Although the associated cut is below the resonance windows considered
here, its real part contributes within those windows. It would therefore
be missed in a dispersion calculation restricted to the pair cut.
We evaluate the neutral channels using the corresponding complete
zero-field radial integral.

By combining the vacuum and thermal contributions, we obtain the full
inverse propagator,
\begin{equation}
 F_m(E)=F_{\rm vac}^{(0)}(E^2)-\Delta\Sigma_m^{\rm vac}(E^2)
        -\delta\Sigma_{m,\rm ch}^T(E)-\delta\Sigma_{\rm neu}^T(E).
 \label{eq:fullinverse}
\end{equation}
For an open Landau pair with longitudinal momentum $k_\ell$, the
vacuum partial widths are
\begin{equation}
 \Gamma_{0,\ell}^{\rm vac}=\frac{g^2\beta k_\ell}{2\pi E^2},
 \qquad
 \Gamma_{m,\ell}^{\rm vac}=\frac{g^2\beta^2r}{2\pi E^2k_\ell}
 \quad(m=\pm1).
\end{equation}
The corresponding gain and loss factors are
$G_\ell=\Gamma_\ell^{\rm vac}n_a n_c$ and
$L_\ell=\Gamma_\ell^{\rm vac}(1+n_a)(1+n_c)$.
Energy conservation, $E_a+E_c=E$, implies the following relations,
which remain valid after summing over the channels:
\begin{equation}
 \frac{G_m}{L_m}=e^{-E/T},\qquad
 \nu_m=L_m-G_m\geq0,\qquad Q_m(E)=E\nu_m(E).
 \label{eq:opticalkms}
\end{equation}
We verify these identities with an independent phase-space calculation.
They test the absorptive part of the self-energy but do not fix the real
matching terms or the prescription for an integrated occupation.

\subsection{Stability in the energy domain of interest}
The positivity of absorption on the real energy rim does not by itself
exclude an unstable zero of $F$ away from the real axis. To check this
possibility, we derive a sufficient bound from the dispersive
representation. In the same real subtraction scheme, we expand the
full vacuum inverse about the spacelike point $s_*=-0.5\,\GeV^2$ as
\begin{equation}
 F_{\rm vac}(s)=F_{\rm vac}(s_*)+Z_{\rm sp}(s-s_*)-R(s),\qquad
 Z_{\rm sp}=\partial_sF_{\rm vac}(s_*).
\end{equation}
For a complex argument $s=x+iy$ with $y>0$, the imaginary part of the
twice-subtracted remainder satisfies
\begin{equation}
 \frac{\Imm R(s)}y=-\frac1\pi\int_{t_{\min}}^\infty\dd t\,
 \widehat Q_{\rm vac}(t)\left[\frac1{|t-s|^2}-\frac1{(t-s_*)^2}\right].
 \label{eq:stabilitydispersion}
\end{equation}
The two terms in brackets must be integrated together: their difference
gives a convergent integral, whereas the separate integrals need not
exist. For $x\geq s_*$, we impose the conditions
\begin{equation}
 (x-t_{\min})^2+y^2\leq(t_{\min}-s_*)^2,\qquad Z_{\rm sp}>0,
 \label{eq:stabilitydomain}
\end{equation}
Under these conditions, the bracket is nonnegative along the entire
cut, and hence $\Imm F_{\rm vac}(s)\geq Z_{\rm sp}y>0$.
The thermal pair numerator and the scattering factor $d(n_c-n_a)$ are
also nonnegative, giving $\Imm\delta\Sigma^T(s)\leq0$.
The bound is unchanged by real energy-independent contacts. For real
$VP$ couplings, the inequality $\Imm[s/(s-m_P^2)]<0$ further
strengthens it. Including energy-dependent local operators would,
however, require a separate stability check.

We use the lowest zero-field pair threshold as a conservative choice of
$t_{\min}$. With this choice, Eq.~\eqref{eq:stabilitydomain} covers
$z=E+i\eta$ with $s=z^2$ and $0<\eta\leq0.1\,\GeV$ throughout the
energy intervals $0.99546\leq E\leq1.04346\,\GeV$ for $\phi$ and
$0.74556\leq E\leq1.04556\,\GeV$ for $K^{*0}$.
For the fields $b=0,1.25,2.5,5,10$, expressed in units of
$10^{-4}\,\GeV^2$, the minimum values of $Z_{\rm sp}$ over all modes
and fields are $0.75799$ and $0.90872$ for the two species, respectively.
The inverse propagator thus has no zeros in these upper-half-plane
rectangles for the temperatures and constant local terms included in
this study. This conclusion establishes stability within a finite
domain; it does not provide a global ultraviolet completion or prove
the full spectral sum rule.

\section{Threshold peaks and spectral integration}
\label{sec:thresholdpeaks}
The Landau-level thresholds require particular care when integrating
the spectral function. To examine their local structure, we consider a
transverse Landau pair with threshold $E_t=u+v$, where $u=\sqrt A$,
$v=\sqrt C$ and $\mu=uv/(u+v)$ is the reduced mass.
Near the threshold, the longitudinal momentum is
$k_z=\sqrt{2\mu(E-E_t)}+\cdots$, and the leading singular part of the
retarded self-energy is
\begin{equation}
 \Sigma_{\rm sing}^R(z)=-\frac{\mathcal C_t}{\sqrt{E_t-z}},\qquad
 \mathcal C_t=\frac{g^2\beta^2r[1+n_B(u)+n_B(v)]}
 {2\pi E_t\sqrt{2\mu}}>0.
 \label{eq:threshold}
\end{equation}
The retarded prescription fixes the square root by continuation from
$z<E_t$ through the upper half-plane. In particular,
$\sqrt{E_t-(E+i0)}=-i\sqrt{E-E_t}$ for $E>E_t$.
The singular term therefore gives
$Q_{\rm sing}=\mathcal C_t/\sqrt{E-E_t}$ above threshold and
$\Ree\Sigma_{\rm sing}=-\mathcal C_t/\sqrt{E_t-E}$ below it.
The absorptive divergence in the open channel and the real divergence
in the closed channel thus follow from the same analytic function.

To determine the spectral structure below threshold, we write the
regular part of the inverse locally as $-D+iQ$, with $D>0$ and $Q>0$.
For $x=E_t-E>0$, the inverse propagator is then approximated by
\begin{equation}
 F(E)\simeq-D+\frac{\mathcal C_t}{\sqrt x}+iQ.
\end{equation}
The real part of this inverse vanishes below threshold at
$x_*=(\mathcal C_t/D)^2$. When $Q\ll D$ and both the regular terms
and neighboring cuts vary slowly over the peak region, we obtain the
following estimates for the halfwidth and integrated area:
\begin{equation}
 \gamma_*\simeq\frac{2Q\mathcal C_t^2}{D^3},\qquad
 \int_{\rm peak}\dd E\,A(E)\simeq\frac{4\pi \mathcal C_t^2}{D^3},\qquad
 \int_{\rm peak}\frac{\dd E}{\pi}\,E A(E)
 \simeq\frac{4E_t\mathcal C_t^2}{D^3}.
 \label{eq:peakarea}
\end{equation}
These area estimates apply to a sufficiently isolated narrow peak and
cannot be used for the entire threshold interval. In particular, the
value of $D$ must be determined from the regular inverse independently
of the position of the zero. To test the estimate, we remove the
singular contribution in Eq.~\eqref{eq:threshold} and extrapolate the
remaining regular part to the threshold.

At the threshold itself, the spectral function vanishes. Its leading
behavior is $A\sim2\sqrt{E-E_t}/\mathcal C_t$ above the endpoint and
$A\sim2Q(E_t-E)/\mathcal C_t^2$ below it, even though the nearby
peak carries a finite area. A positive set of spectral samples can
therefore miss part of this area. Checking the absorptive identity at
every grid point does not resolve the problem either. This
nonuniform behavior is particularly consequential for small differences
between spin moments. Accordingly, we do not assume a uniform $b^2$
expansion on the unsmeared real energy axis.

\begin{table}[htbp]\centering\small
\caption{Closed transverse $\phi$ thresholds at $T=0.15\,\GeV$ and
$b=2.5\times10^{-4}\,\GeV^2$. The last column compares the leading
weighted-area estimate in Eq.~\eqref{eq:peakarea} with the local residue
$2E_*/\partial_E\Ree F(E_*)$. This comparison does not test the integral
over a full threshold interval.}
\label{tab:threshold}
\begin{tabular}{rrrrrr}\toprule
$r$ & $E_t$ (GeV) & $E_t-E_*$ (GeV)& $\gamma_*$ (GeV)&$Z_{\rm peak}$& Rel. residue difference\\\midrule
13 & 1.000433742 & $8.0715\times10^{-9}$ & $2.1090\times10^{-10}$ & $7.2482\times10^{-7}$ & $4.5856\times10^{-6}$\\
17 & 1.004424051 & $2.0325\times10^{-8}$ & $1.1521\times10^{-9}$ & $2.2391\times10^{-6}$ & $9.7569\times10^{-6}$\\
21 & 1.008398570 & $5.0442\times10^{-8}$ & $5.5790\times10^{-9}$ & $7.1636\times10^{-6}$ & $1.7815\times10^{-5}$\\
\bottomrule\end{tabular}\end{table}

\begin{figure}[htbp]
 \centering\figfile{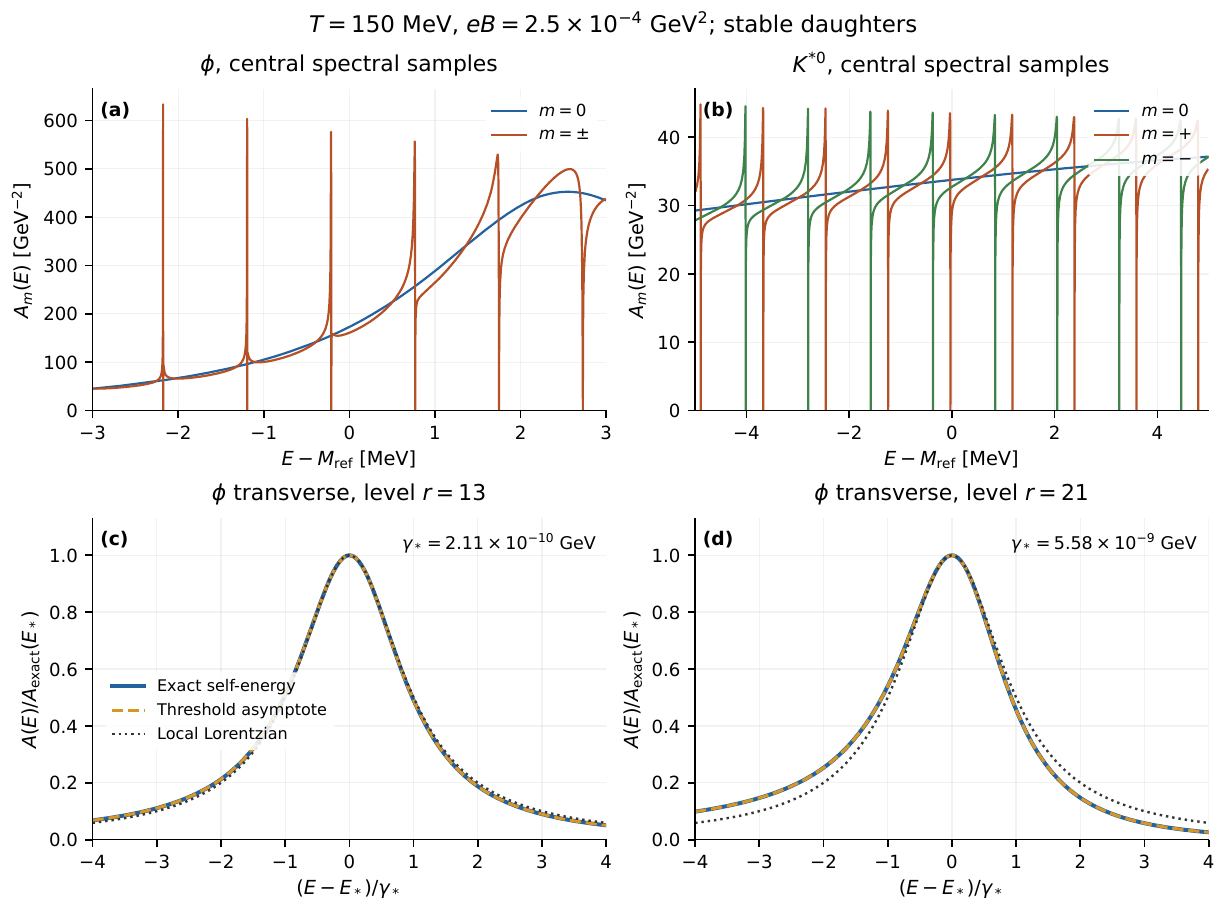}{1}
 \caption{Spectral functions on the real energy rim at the central
temperature and magnetic field. The upper panels show samples with the
threshold positions taken into account. The curves connecting these
samples do not establish convergence of the narrow-peak areas, which
are evaluated by a separate adaptive integration. The lower panels
resolve two transverse $\phi$ peaks and compare the result from the
exact self-energy with the independently extracted threshold asymptote
and local Lorentzian. All three curves are divided by the same
$A_{\rm exact}(E_*)$. The scale $\gamma_*$ denotes the halfwidth refined
using the derivative. No daughter width or imaginary energy resolution
is included.}
 \label{fig:threshold}
\end{figure}

\section{Daughter spectral widths and threshold scales}
\label{sec:width}
We now investigate the daughter spectral scale that can modify the
narrow peaks described by Eq.~\eqref{eq:peakarea}. Finite daughter
damping in vector spectra has been considered previously, including
self-consistent pion dressing and the limitations associated with
current vertices \cite{VanHeesKnoll2001,VanHeesKnoll2002}. Here, we
restrict the calculation to the leading isolated threshold and match
its regular coefficients $D,Q$ to the stable-daughter result. This
construction allows us to examine the local threshold response, but
does not replace the full matched loop or determine an in-medium
kaon collision rate.

The local regime considered here differs from a pole--threshold
coincidence. In the analysis of Patk\'os, Sz\'ep and Sz\'epfalusy
\cite{Patkos2003}, the inverse propagator is expanded in nonnegative
powers of a square-root threshold coordinate, and its regular term
is tuned to zero. In the present transverse Landau channel, the
self-energy instead has an inverse-square-root singularity at nonzero
$D$. The resulting detuning scale therefore characterizes a different
regime. A pole--threshold coincidence requires a separate analysis.

\subsection{A positive two-daughter spectral construction}
To introduce daughter widths, we start from the equilibrium bosonic
spectral representation. The retarded loop contains the kernel
\begin{equation}
 \int\frac{\dd\omega_a\dd\omega_c}{(2\pi)^2}
 \frac{\rho_a(\omega_a)\rho_c(\omega_c)
 [1+n_B(\omega_a)+n_B(\omega_c)]}
 {E-\omega_a-\omega_c+i0},
 \label{eq:doublespectral}
\end{equation}
where the momentum measure and vertex are specified separately.
Both signs of the frequencies must be retained to obtain the full
pair and scattering loop \cite{Weldon1983}. With positive spectral
inputs, the real and imaginary parts belong to the same causal
function. The spectral shapes and their tails nevertheless remain
model inputs, as also discussed for vacuum spectral convolutions
\cite{CrivellinHoferichter2023}.

Near the selected positive-frequency pair endpoint, we assign each
daughter an independent, normalized energy displacement
$p_h(\delta)=\Theta(h-|\delta|)/(2h)$ about its stable dispersion.
The positive-frequency support remains positive provided that
$h<\min(u,v)$. At fixed momentum, the daughter spectrum is defined by
$\rho_h(\omega>0;k)=\pi p_h(\omega-\epsilon(k))/\omega$, with
$\rho_h(-\omega;k)=-\rho_h(\omega;k)$ and
$\int_0^\infty\dd\omega\,\omega\rho_h/\pi=1$.
We hold the threshold vertex, quasiparticle residue, and Bose enhancement
at their central values. The convolution of the two daughter
displacements then gives the pair distribution
\begin{equation}
 W_h(\delta)=\frac{2h-|\delta|}{4h^2}
 \Theta(2h-|\delta|),\qquad \int\dd\delta\,W_h=1.
\end{equation}
After integrating over the longitudinal threshold momentum and the
two daughter spectra, we obtain
\begin{align}
 \Sigma_h^R(z)&=-\mathcal C_t K_h(E_t-z),\qquad
 F_h=-D+iQ+\mathcal C_tK_h,\\
 K_h(a)&=\int\frac{\dd\delta\,W_h(\delta)}{\sqrt{a+\delta}}
 =\frac{(a+2h)^{3/2}-2a^{3/2}+(a-2h)^{3/2}}{3h^2}.
 \label{eq:boxkernel}
\end{align}
On the real rim, the argument is $a=x-i0$, and the same retarded
branch is used for every power. It follows that $\Imm K_h\geq0$
and, for $Q>0$, $A_h\geq0$. The stable-daughter kernel
$1/\sqrt a$ is recovered in the limit $h\to0$.
As a numerical check, direct integration of the pair distribution
agrees with the closed expression to a relative accuracy of
$9.2\times10^{-15}$ at the tested points. The subtraction terms,
contact contribution, and other thresholds are retained in the
matched regular background. Dressing their full energy dependence
and the associated vertices requires a separate calculation.

To identify the scale that controls the response to daughter widths,
we introduce $t=x/x_*$, $q=Q/D$, $H=h/x_*$ and
$\mathcal K_H(t)=\sqrt{x_*}K_h(x_*t-i0)$. In these variables, the
local spectrum takes the form
\begin{equation}
 \frac{D A_h}2=
 \frac{q+\Imm\mathcal K_H(t)}
 {[-1+\Ree\mathcal K_H(t)]^2+
 [q+\Imm\mathcal K_H(t)]^2}.
 \label{eq:scaledwidth}
\end{equation}
Thus, the daughter-width dependence is controlled by $h/x_*$ rather
than $h/M$. By maximizing the real part of the kernel independently,
we find
\begin{equation}
 \max_x\Ree K_h=\frac{4\sqrt6}{9\sqrt h},\qquad
 x_{\max}=\frac{2h}{3},\qquad
 h_c=\frac{32}{27}x_*.
 \label{eq:boxcritical}
\end{equation}
The real zeros of the local $\Ree F_h$ merge at equality and are
absent for $h>h_c$. This criterion concerns the real zeros only.
It does not imply that a continued complex pole disappears or that
all integrated threshold weight is lost.

For comparison, we also consider narrow Cauchy quasiparticle peaks
with individual energy halfwidths $\gamma_a,\gamma_c$. Their
convolution has the pair halfwidth $\eta=\gamma_a+\gamma_c$ and gives
the local expressions
\begin{equation}
 K_\eta(x)=(x-i\eta)^{-1/2},\qquad
 \eta_c=\frac{3\sqrt3}{8}x_*.
 \label{eq:cauchycritical}
\end{equation}
The Cauchy tails are used here as a local quasiparticle approximation;
they do not define globally positive-energy daughter spectra. The
kernel in Eq.~\eqref{eq:cauchycritical} follows from the convolution
of the two peaks and does not justify the replacement
$E\mapsto E+i\eta$ in the complete external propagator.
For $\eta\ll x_*$, the peak halfwidth is $\gamma_*+\eta$ to leading
order. Its height and width can therefore change at
$\eta\sim\gamma_*$, before the real zeros merge at
$\eta\sim x_*$.

The compact box spectrum has a different response at small widths
because it produces no daughter-induced absorption below its lowest
pair support. In the zero-$Q$ limit, its root first enters that support
at $h/x_*=[(8-4\sqrt2)/3]^2\simeq0.6100$.
Consequently, the early change in the linewidth depends on the
spectral tails. The common parametric detuning scale is more robust
than the numerical coefficient, which depends on the spectral shape.

\begin{table}[htbp]
\centering\small
\caption{Sensitivity scales for three central $\phi$ thresholds, in eV.
Here $h$ is the half-support of an individual box daughter spectrum,
whereas $\eta$ is the pair halfwidth for Cauchy spectra. Neither
parameter represents a predicted collision rate.}
\label{tab:widthscales}
\begin{tabular}{rrrrr}\toprule
$r$ & $x_*$ & $\gamma_*$ & $\eta_c$ & $h_c$\\\midrule
13 & 8.0715 & 0.21090 & 5.2426 & 9.5662\\
17 & 20.325 & 1.1522 & 13.202 & 24.089\\
21 & 50.442 & 5.5791 & 32.763 & 59.784\\\bottomrule
\end{tabular}
\end{table}
The corresponding scales are summarized in Table~\ref{tab:widthscales}.
Since neighboring pair thresholds are separated by approximately
$1\,\MeV$, the hierarchy $x_*,h\ll\Delta E_{\rm LL}$ is well
satisfied when the real zeros merge. We also check the approximation
of a frozen thermal prefactor by restoring the displaced thermal
enhancement and scalar residues in the two-dimensional box integral.
For the tested $h\leq100\,\mathrm{eV}$, the relative change in the
complex kernel is less than $9.0\times10^{-8}$. This check constrains
the frozen prefactor, but does not test the uncomputed daughter
interactions or vertex corrections. Scans reaching
$h\sim\Delta E_{\rm LL}$ are formal diagnostics outside the
isolated-threshold approximation.

\begin{figure}[htbp]
 \centering\figfile{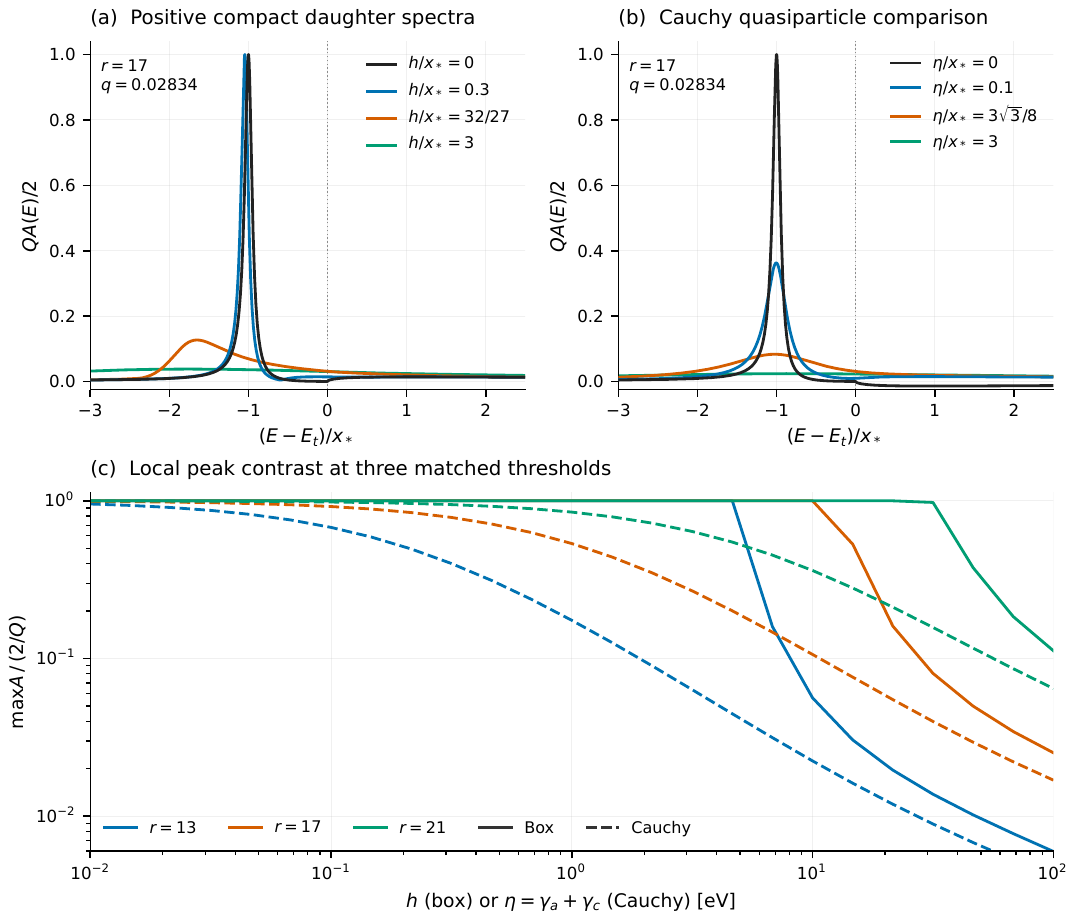}{1}
 \caption{Local threshold spectra matched to the full stable-daughter $\phi$ calculation at $T=150$ MeV and $b=2.5\times10^{-4}$ GeV$^2$. Panels (a,b) show the results for $r=17$ and $q=Q/D=0.02834$. Here $h$ denotes the half-support of each compact daughter spectrum, while $\eta=\gamma_a+\gamma_c$ is the pair Cauchy HWHM. The orange curves correspond to the respective values at which the real zeros merge; these values do not give universal criteria for peak disappearance. Panel (c) shows the maximum contrast in $|E-E_t|<100x_*$ for three thresholds and widths up to 100 eV. The coefficients $C,D,Q$ are held fixed. These results describe a local causal convolution of daughter spectra, rather than a full dressed-loop spectrum or a calculated collision width.}
 \label{fig:finitewidth}
\end{figure}

\subsection{Peak areas and spectral redistribution}
We first check the stable-daughter peak area by independently
integrating the \emph{complete} matched inverse. This tests the area
formula without relying on its local derivative. Let $E_r$ denote
the resolved zero and $x_r=E_t-E_r$ its distance from the threshold.
Below the selected threshold, we define
$F_{\rm reg}=F-\mathcal C_t/\sqrt{E_t-E}$ and
$A_{\rm reg}=-2\Imm F_{\rm reg}^{-1}$, and write
$Z_*=4E_tx_*/D$ for the leading weighted narrow-peak area.
The integral over the peak core is
\begin{equation}
 Z_{\rm core}(W)=\int_{E_r-W}^{E_r+W}
 \frac{\dd E}{\pi}E[A(E)-A_{\rm reg}(E)],\qquad W=0.4x_r.
\end{equation}
For $r=13,17,21$, the resulting values are
$(0.694948,2.035716,5.885488)\times10^{-6}$, respectively.
We compare them with Eq.~\eqref{eq:peakarea} after multiplying by
the retained Lorentzian fraction $(2/\pi)\arctan(W/\gamma_*)$.
The deviations are $+0.0325\%,-0.1338\%,-0.8089\%$.
Increasing the quadrature from 64 to 128 points in
$E=E_r+\gamma_*\tan\theta$ changes the integrals by at most
$5.2\times10^{-8}$ relatively. The asymptotic area is therefore
verified at about the percent level for these finite windows.
The much smaller residue differences in Table~\ref{tab:threshold}
refer to a different test. Varying $W/x_r$ from 0.1 to 0.8 checks
the specified convention for the tails; it does not define a unique
observable separation between the peak and its background.

To examine redistribution in the finite-width \emph{local} model,
we compare the spectrum before and after dressing within the same
fixed patch. We define the changes on the two sides of the threshold by
\begin{equation}
 \Delta Z_h^{\lessgtr}(L)=\int_{I_{\lessgtr}}
 \frac{\dd E}{\pi}E[A_h(E)-A_0(E)],\quad
 I_<=[E_t-L,E_t],\quad I_>=[E_t,E_t+L].
 \label{eq:transfer}
\end{equation}
The chosen patch half-widths are
$L=\Delta E_{\rm LL}/32,\Delta E_{\rm LL}/16,
\Delta E_{\rm LL}/8$, all much larger than $x_*$.
For $r=17$, $h=100\,\mathrm{eV}$ and $L=\Delta E_{\rm LL}/16$,
the box spectrum gives
$(\Delta Z_h^<,\Delta Z_h^>)/Z_*=(-0.249761,+0.249757)$,
even though $h>h_c$. The two contributions almost cancel, leaving
a net patch change of $-4.0\times10^{-6}Z_*$.
Thus, the weight is transferred from below to above the unshifted
threshold instead of simply being removed. For the Cauchy spectrum
at $\eta=100\,\mathrm{eV}$, the corresponding changes are
$(-0.155946,+0.140515)Z_*$, with a larger leakage out of the patch.

\begin{figure}[htbp]
 \centering\figfile{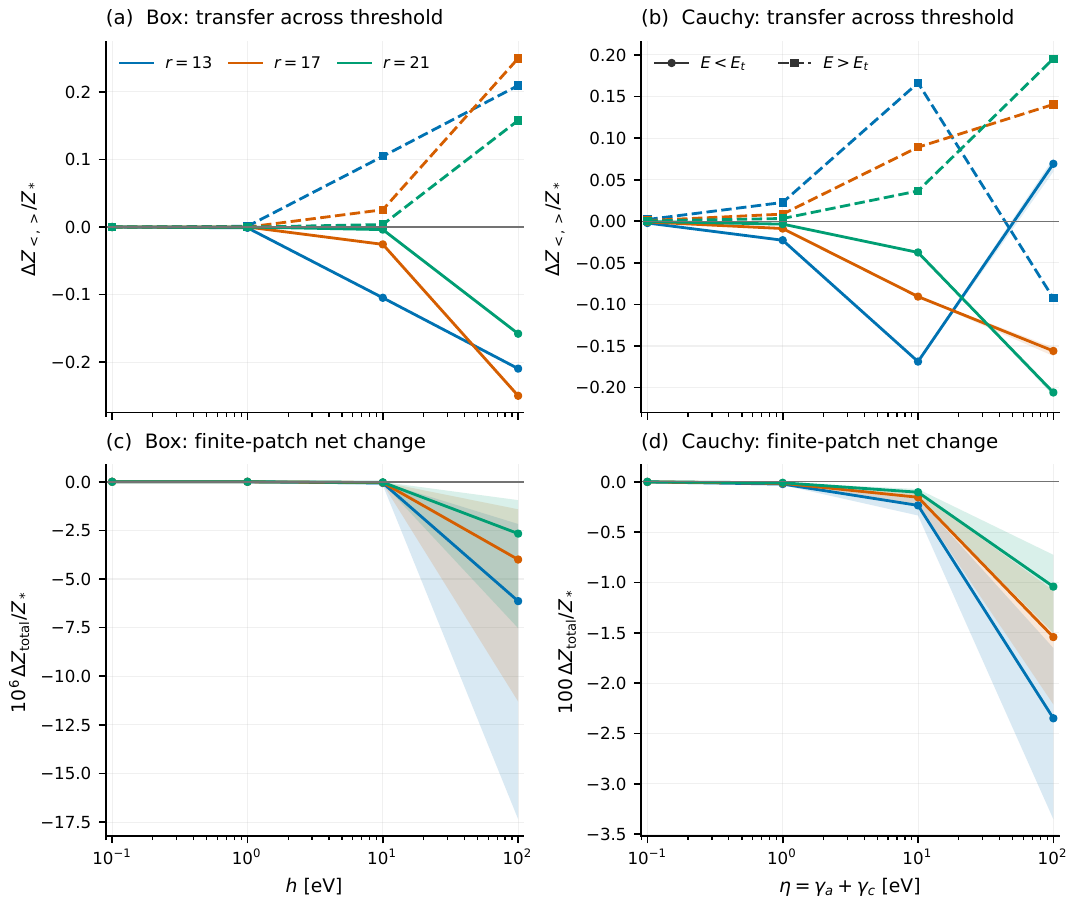}{1}
 \caption{Redistribution across the nominal threshold in the local matched model. The quantity $\Delta Z$ is the integral of $E[A_h(E)-A_0(E)]/\pi$, normalized by the stable narrow-peak residue $Z_*$; $<$ and $>$ label the two sides of $E_t$. The central curves use the patch half-width $L=\Delta E_{\rm LL}/16$, while the shaded ranges correspond to varying $L$ from $\Delta E_{\rm LL}/32$ to $\Delta E_{\rm LL}/8$. Panels (a,b) show the separate contributions, and panels (c,d) show their sum on the different scales indicated. The bands represent patch dependence, rather than statistical errors; connecting lines guide the eye. These finite-patch integrals are not spectral sum rules. The compact and Cauchy widths differ in both their definitions and their tail behavior.}
 \label{fig:transfer}
\end{figure}

Figure~\ref{fig:transfer} shows how these results depend on the
chosen patch and the spectral tails. Such dependence is part of the
finite-patch calculation. The balances do not constitute global sum
rules or finite-width predictions for $r_{00}$. The connection to
weighted moments and a specified endpoint intervention in the full
background is given in Sec.~\ref{sec:widthbridge}.

\subsection{Validity of the threshold resummation}
At the generated root, $\mathcal C_t/\sqrt{x_*}=D$, and repeated
insertions of the retained parent self-energy must be Dyson summed.
A daughter energy shift or damping changes the singular term by a
relative amount of order $(|\delta E_t|+\eta)/(2x_*)$.
Summing the parent self-energy alone therefore does not suppress
this separate enhancement. The kernels introduced above resum one
specified contribution of this kind. As in other consistent resonance
expansions, control requires a hierarchy of scales in addition to a
small coupling \cite{Beneke2015}.

For an omitted regular correction $\delta F=\delta R+i\delta Q$,
sufficient conditions for local control of the slope and residue are
\begin{equation}
 |\delta R|\ll D,\quad
 |\partial_E\delta R|\ll D/(2x_*),\quad
 |\delta\mathcal C_t|\ll\mathcal C_t.
\end{equation}
To preserve the original linewidth, one also needs
$|\delta Q|\ll Q$. Keeping the pointwise location within that width
requires the stronger condition $|\delta R|\ll Q$.
Short-range interactions that are irreducible in the parent channel
can receive an enhancement from the one-dimensional pair loop
$G_{1d}\sim-\mu/\kappa$, $\kappa=\sqrt{2\mu x_*}$.
Their perturbative treatment requires
$|\lambda_{1d}|\mu/\kappa\ll1$, where $\lambda_{1d}$ is a
separately matched interaction. The present inputs do not establish
this inequality. Consistent vertices are also required when dressing
a conserved-current response \cite{VanHeesKnoll2002}.

These conditions delimit the stable-daughter benchmark; they do not
establish an all-order hadronic prediction. At fixed nonzero daughter
width, the real-zero criterion eventually fails as
$\mathcal C_t\to0$. At zero width, however, the idealized root
exists for every $\mathcal C_t>0$. The joint limit is therefore
nonuniform for the scaled profile and root count. This does not
establish noncommuting limits for the absolute integrated weight,
since $Z_*\propto\mathcal C_t^2$ can vanish along both paths.
Similarly, a smooth moment over the full window can have a regular
weak-field limit despite the nonuniformity of each resolved local
profile.

The magnetic-field dependence also distinguishes the scale of a
narrow peak from the separation of neighboring thresholds.
For a fixed Landau index $r$, Eq.~\eqref{eq:threshold} gives
$\mathcal C_t\propto b^2$ and hence $x_*\propto b^4$, provided
that $D$ has a finite nonzero limit. For a sequence of thresholds at
fixed energy above the zero-field pair threshold, $r b$ instead
approaches a positive constant. In this case,
$\mathcal C_t\propto b$ and $x_*\propto b^2$.
For either sequence,
$\Delta E_{\rm LL}\simeq b(1/u+1/v)\propto b$.
Away from a parent-pole or an additional threshold degeneracy,
these scalings allow the hierarchy
\begin{equation}
 x_*\ll h\ll\Delta E_{\rm LL}.
 \label{eq:twoscales}
\end{equation}
In this regime, the narrow subthreshold real zeros can already be
absent while neighboring Landau thresholds remain resolved.
This local result does not determine the weak-field power of the
integrated spin moment. Moreover, the scaling of
$\gamma_*=2Qx_*/D$ depends on $Q$ and cannot be assumed to coincide
with that of $x_*$ when other open channels close along the limiting
sequence.

\section{Spectral moments and the operator convention}
\label{sec:moments}
To characterize the spin dependence directly from the calculated
spectra, we define the energy-weighted moments and their ratio by
\begin{equation}
 N_m^{I,E}=\int_I\frac{\dd E}{\pi}\,E A_m(E)n_B(E),\qquad
 Z_m^I=\int_I\frac{\dd E}{\pi}\,E A_m(E),\qquad
 r_{00}^{I,E}=\frac{N_0^{I,E}}{\sum_mN_m^{I,E}}.
 \label{eq:momentE}
\end{equation}
The notation $r_{00}$ emphasizes that this is a model spectral ratio.
An energy-weighted spectral prescription is also used in
Ref.~\cite{ShengMagnetic2024}. A related Wigner construction places
a fixed on-shell energy outside the frequency integral and explicitly
identifies the result with occupation beyond leading order
\cite{DongLinearResponse2024}. At rest and for a fixed reference
mass, this motivates comparison with the alternative prescription
\begin{equation}
 N_m^{I,M}=\int_I\frac{\dd E}{\pi}\,M A_m(E)n_B(E),\qquad
 r_{00}^{I,M}=\frac{N_0^{I,M}}{\sum_mN_m^{I,M}}.
 \label{eq:momentM}
\end{equation}
The two measures agree for a common free pole at $E=M$.
For a broad interacting spectrum, however, KMS does not make them
equivalent. Neither prescription separately normalizes the spin
spectra within the finite window.

The choice of weight does not remove the dependence on the operator
convention. To illustrate this point, consider a local invertible
field change $V'_m=(1+\alpha_m b^2)V_m$,
$\alpha_+=\alpha_-\equiv\alpha_T$. The bare-field moments then
transform as $N'_m=(1+\alpha_m b^2)^2N_m$.
Writing $r=r_{00}^{I,E}$ at the same finite field, we obtain the
exact relation
\begin{equation}
 r'=\frac{r(1+\alpha_0b^2)^2}
 {r(1+\alpha_0b^2)^2+(1-r)(1+\alpha_Tb^2)^2}.
\end{equation}
Expanding only the imposed rescaling, with $[\alpha_m]=\GeV^{-4}$,
gives
\begin{equation}
 r_{00}'-r_{00}=2b^2r_{00}(1-r_{00})(\alpha_0-\alpha_T)+O(b^4).
 \label{eq:fieldredefinition}
\end{equation}
If $r_{00}\to1/3$ as $b\to0$, the leading coefficient becomes
$4(\alpha_0-\alpha_T)/9$, with an $o(b^2)$ remainder.
This statement does not require a uniform analytic expansion of the
unsmeared spectrum.

The change in the field moment does not represent a change in physics.
For a matched physical source overlap $g_mV_m$, the transformation
$g'_m=g_m/(1+\alpha_m b^2)$ leaves $|g_m|^2A_m$ invariant.
This is an application of standard field-redefinition invariance
\cite{Criado2019}. It shows why an interpolating-field moment cannot
be interpreted as an observable without specifying the source.
In the present benchmark, we fix a canonical field basis and use
Eq.~\eqref{eq:momentE}; this convention does not establish a
basis-independent meson number.

A further distinction is needed when relating the spectral ratio to
a decay measurement. For an asymptotic spin-one ensemble decaying
into two spinless particles, the familiar polar-angle distribution is
$\dd N/\dd\cos\theta=\tfrac34[(1-\rho_{00})+(3\rho_{00}-1)\cos^2\theta]$.
A reconstructed in-medium sample also depends on production, decay
channels, nonresonant contributions, daughter transport, and mass
selection. In particular, charged Landau wavefunctions do not
automatically satisfy the vacuum angular template.
Equations~\eqref{eq:momentE} and \eqref{eq:momentM} alone do not
provide a description of such a sample.

In the construction of Ref.~\cite{ZhuShengHou2025}, a thermal source
correlator and vacuum propagation are contracted with the kaon decay
vertex. The invariant-mass average includes both decay kinematics
and vertex factors, while a kaon loop with pole kaon propagators
generates the parent $\phi$ width. Our field moments and selected
replacements of daughter thresholds describe a different operation.
They do not provide the corresponding production and decay matching.

To examine how cancellations enter the ratio, it is useful to write
an exact identity valid for any fixed positive weight $w(E)$:
\begin{equation}
 r_{00}^I[w]-\frac13=
 \frac{2\int_I\dd E\,w(E)
 [A_0-(A_++A_-)/2]}{3\int_I\dd E\,w(E)(A_0+A_++A_-)}.
 \label{eq:cancellation}
\end{equation}
The deviation of the spin ratio is a signed tensor area divided by
a positive total area. Positivity of $A_m$ therefore does not, by
itself, determine the sensitivity to the window boundaries; that
sensitivity must be examined explicitly.
At the central parameters, the sum of the absolute tensor integrals
over individual threshold intervals, divided by the absolute net
integral, is $6913$ for $\phi$ and $9038$ for $K^{*0}$.
These values give lower bounds on the total cancellation ratio,
because further cancellations can occur within each interval.
The maximum accumulated signed area is $477$ and $56$ times the
final net magnitude, respectively. For this cumulative diagnostic,
the denominator is fixed to its full-window value. The intermediate
points therefore do not represent ratios evaluated in successively
smaller windows.
\section{Numerical results}
\label{sec:numerical}
\subsection{Numerical implementation and convergence}
We evaluate the vacuum remainder directly on the real rim and use an
analytic pole subtraction for the thermal integral. In particular, the
identity
\begin{equation}
 \frac1{(E_a+E_c)^2-s}
 =\frac{s-(E_a-E_c)^2}{4s(k_z^2-k_0^2)},\qquad
 k_0^2=\frac{\lambda(s,A,C)}{4s},
\end{equation}
allows us to integrate the singular part as a logarithm for an open pair
and an arctangent for a closed pair. We accelerate the contribution of
distant, smooth thermal levels with convergent moments. To interpolate
the remaining energy dependence, we first extract each closed-channel
square-root term and store only its analytic remainder. The branch
terms and pair cuts are therefore evaluated exactly throughout the
calculation.

We examine the self-energy by varying the vacuum matching point,
approaching the real rim from the upper half-plane, refining the thermal
cutoff and quadrature, and taking the smooth spacelike zero-field limit.
At the central parameters, the maximum relative mismatch on the spectral
grids is $6.9\times10^{-15}$ for the optical identity and
$1.4\times10^{-15}$ for the KMS ratio. In the independent thermal
calculation, increasing the cutoff from $24T$ to $28T$ changes the
self-energy by at most $1.1\times10^{-11}\,\GeV^2$. These comparisons
test the implementation of the specified loop; they do not estimate
the effects of omitted interactions.

The narrow threshold peaks require a separate treatment in the energy
integral. We divide the integration range at every Landau threshold
and locate the zeros of $\Ree F_m$ with logarithmic probes in each
interval. Additional panels resolve fractions and multiples of the
local width $\Imm F_m/|\partial_E\Ree F_m|$, after which we apply
adaptive integration with a squared-sine endpoint map. We exclude
numerical strips of width $2\times10^{-13}\,\GeV$ at the endpoints
and bound their omitted weight with $A_m\leq2/Q_{\rm neutral}^{\rm vac}$.
This exclusion is a numerical device and introduces no physical
regularization.

At the central parameters, independently refining the root search and
adaptive tolerance changes $r_{00}$ by less than $10^{-12}$.
We nevertheless retain the larger endpoint-strip bounds in the quoted
accuracy. The importance of resolving the peaks can be seen from a
fixed-order Gauss calculation: the 24- and 48-point results for $\phi$
are $0.33333773$ and $0.33332373$, respectively, whereas the resolved
integral gives $0.33333535$. The intermediate result would imply the
wrong sign even though the spectrum satisfies the pointwise optical
identity.

\subsection{Comparison with published results in common limits}
We also compare with independent evaluations of published expressions
in limits shared by the respective models. For the scalar-current
vacuum self-energy, Eqs.~(B.4)--(B.6) of
Ref.~\cite{BandyopadhyayMallik2017} agree with our result after their
spatial trace is divided by three and the same two real subtractions
are applied. For the longitudinal absorptive part at $\bm p=0$, we
find agreement with Eqs.~(52) and (E5) of Ref.~\cite{RhoGeneral2019}
after using the energy-dependent vertex relation $g=g_R E^2/2$,
$g_R=20.72\,\GeV^{-2}$. In this comparison, the reference calculation
explicitly sums the energy-conservation roots and their delta-function
Jacobians, independently of our self-energy routine.

\begin{table}[htbp]\centering\small
\caption{Comparison with published expressions in shared limits.
The relative differences characterize numerical agreement between
formulas and do not measure physical accuracy.}
\label{tab:published}
\begin{tabular}{lrr}\toprule
Quantity & Points & Maximum relative difference\\\midrule
Twice-subtracted vacuum self-energy & 6 & $8.11\times10^{-15}$\\
Vacuum decay width & 6 & $1.78\times10^{-15}$\\
Mapped thermomagnetic longitudinal cut & 40 & $7.33\times10^{-15}$\\
Direct full thermal self-energy cut & 4 & $8.89\times10^{-15}$\\\bottomrule
\end{tabular}\end{table}
The agreement in these limits does not make the underlying effective
actions equivalent. Ref.~\cite{BandyopadhyayMallik2017} includes the
magnetic field in the propagators but retains an ordinary-derivative
strong vertex. The transverse Landau weights in
Ref.~\cite{RhoGeneral2019} likewise differ from those of the
current-connection action used here. The comparisons therefore do not
amount to reproducing either complete finite-field spectrum.

\subsection{Dependence on field, temperature and energy window}
We first consider the dependence of the spectral ratio on the integration
window. At $b=2.5\times10^{-4}\,\GeV^2$ and $T=150\,\MeV$,
the $\phi$ ratio for half-windows $w=10,20,24\,\MeV$ is
$r_E-1/3=(-37.588,2.022,12.618)\times10^{-6}$.
For $K^{*0}$, the corresponding results at $w=50,100,150\,\MeV$ are
$(6.019,-0.516,-0.215)\times10^{-6}$.
The changes are larger than the numerical integration indicators in
Table~\ref{tab:scan}, and the sign depends on the window for both
species. To check these reversals, we increase the root-search density
and tighten the adaptive tolerances for the narrow $\phi$ window and
the point at $b=5\times10^{-4}\,\GeV^2$. Either weighting prescription
changes by at most $1.1\times10^{-14}$, leaving the sign reversals
unchanged.

At the central window, the four-field $\phi$ scan gives
$(-2.212,2.022,-30.260,-44.437)\times10^{-6}$ for
$b=(1.25,2.5,5,10)\times10^{-4}\,\GeV^2$.
This sequence does not support a uniform positive quadratic dependence
of the sharp-window moment evaluated on the real rim. The temperature
dependence is smoother: for $\phi$ at $T=120,150,170\,\MeV$ and the
central field and window, the deviation remains positive and takes the
values $(1.515,2.022,2.206)\times10^{-6}$.
Thus a numerically converged spectral ratio can still depend strongly
on the energy projection used to define it.

\begin{table}[htbp]\centering\small
\caption{Results of the primary loop scan with $\xi=0$.
The integration interval is $I=[M-w,M+w]$, and the central parameters are
$T=150$ MeV, $b=2.5\times10^{-4}\,\GeV^2$, $w_\phi=20$ MeV, and
$w_{K^*}=100$ MeV. Deviations and numerical indicators are given in parts
per million. The indicator $\epsilon_{\rm num}$ combines the
$E$-weighted quadrature estimate with an omitted-strip bound and excludes
model and parameter uncertainties.}
\label{tab:scan}
\begin{tabular}{lrrrrrr}\toprule
$V$ & $T$ (MeV)& $10^4b/\GeV^2$ & $w$ (MeV)& $10^6(r_E-1/3)$ & $10^6(r_M-1/3)$ & $10^6\epsilon_{\rm num}$\\\midrule
$\phi$ & 150 & 2.5 & 10 & -37.5878 & -37.2412 & 0.0078\\
$\phi$ & 120 & 2.5 & 20 & +1.5150 & +1.4329 & 0.0237\\
$\phi$ & 150 & 1.25 & 20 & -2.2117 & -2.1804 & 0.0466\\
$\phi$ & 150 & 2.5 & 20 & +2.0216 & +1.9402 & 0.0238\\
$\phi$ & 150 & 5 & 20 & -30.2602 & -29.9821 & 0.0124\\
$\phi$ & 150 & 10 & 20 & -44.4372 & -44.4143 & 0.0066\\
$\phi$ & 170 & 2.5 & 20 & +2.2064 & +2.1315 & 0.0245\\
$\phi$ & 150 & 2.5 & 24 & +12.6179 & +12.3496 & 0.1672\\
$K^{*0}$ & 150 & 2.5 & 50 & +6.0193 & +5.9867 & 0.0050\\
$K^{*0}$ & 120 & 2.5 & 100 & -0.3267 & -0.3170 & 0.0094\\
$K^{*0}$ & 150 & 1.25 & 100 & -0.4027 & -0.3669 & 0.0183\\
$K^{*0}$ & 150 & 2.5 & 100 & -0.5164 & -0.4860 & 0.0093\\
$K^{*0}$ & 150 & 5 & 100 & +5.7731 & +5.7968 & 0.0050\\
$K^{*0}$ & 150 & 10 & 100 & +4.7639 & +7.1748 & 0.0027\\
$K^{*0}$ & 170 & 2.5 & 100 & -0.7496 & -0.6981 & 0.0094\\
$K^{*0}$ & 150 & 2.5 & 150 & -0.2151 & -0.1462 & 0.0173\\
\bottomrule\end{tabular}\end{table}

\begin{figure}[htbp]
 \centering\figfile{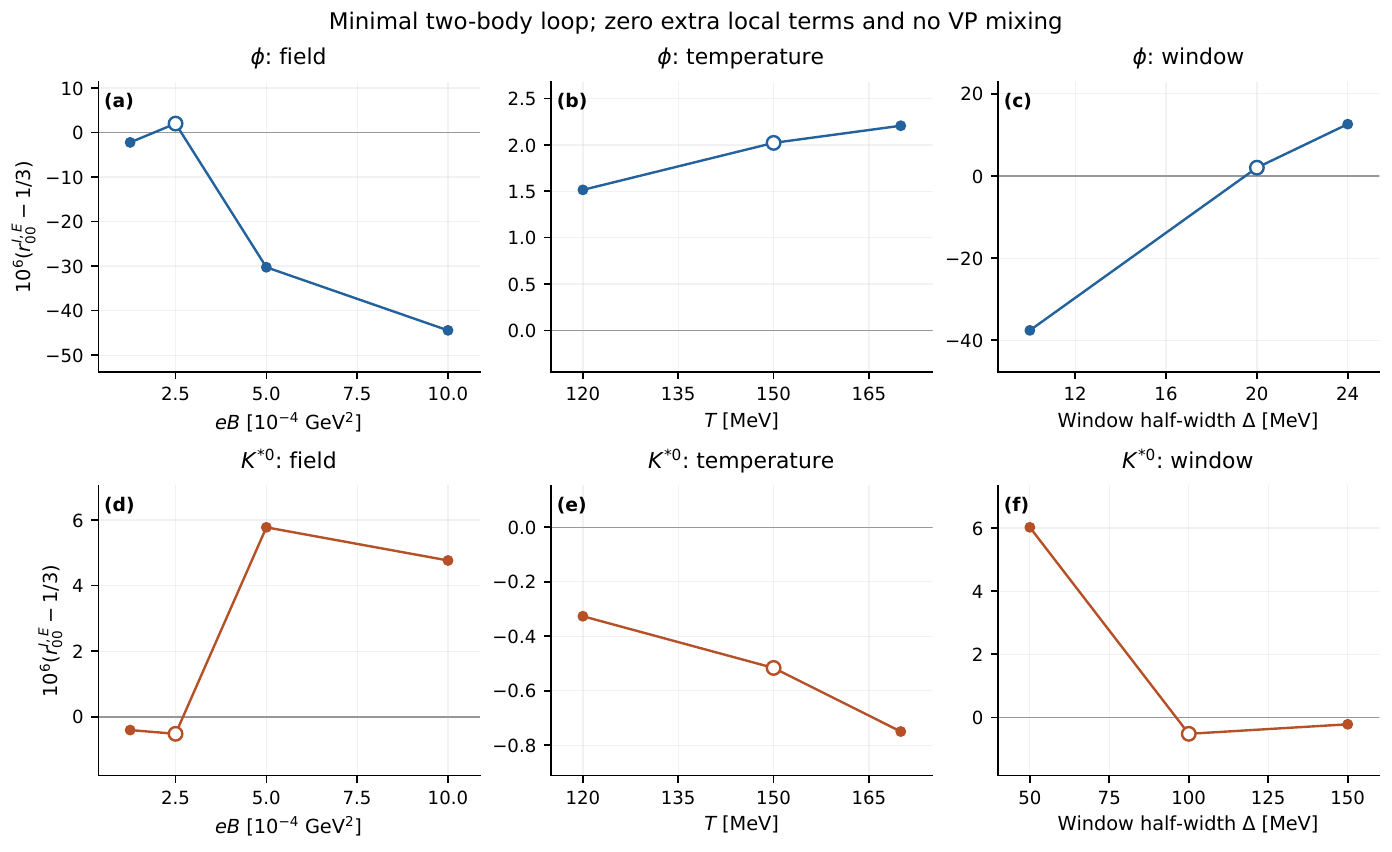}{1}
 \caption{Spectral ratio in the minimal two-body model as a function of
field, temperature and symmetric energy window. Each panel varies one
input while the others remain fixed. Additional local electromagnetic
coefficients and $VP$ mixing are omitted. Open circles indicate the
central configuration. The connecting segments guide the eye and are
neither interpolations nor weak-field fits. Error bars combine the
integration estimate and the bound on omitted endpoint strips; they can
be smaller than the symbols and contain no model uncertainty.}
 \label{fig:scans}
\end{figure}

\begin{table}[htbp]\centering\small
\caption{Comparison of projection prescriptions and local terms at the
central parameters. The derivative is calculated from
$\xi=\pm1\,\GeV^{-3}$. The coefficient $\xi_{\rm lin}$ denotes the
linearized zero of $r_E-1/3$. All its values lie outside the sampled
range and are extrapolations, rather than determined physical
coefficients.}
\label{tab:sensitivity}
\begin{tabular}{llrrrr}\toprule
$V$ & Kernel & $10^6(r_E-1/3)$ & $10^6(r_M-1/3)$ & $10^7\partial_\xi r_E/\GeV^3$ & $\xi_{\rm lin}$ ($\GeV^{-3}$)\\\midrule
$\phi$ & Loop & +2.0216 & +1.9402 & -1.91762 & +10.5420\\
$\phi$ & Loop+$VP$ & +1.2033 & +1.0925 & -1.91762 & +6.2749\\
$K^{*0}$ & Loop & -0.5164 & -0.4860 & -1.01625 & -5.0811\\
$K^{*0}$ & Loop+$VP$ & -0.6612 & -0.6460 & -1.01625 & -6.5065\\
\bottomrule\end{tabular}\end{table}

\begin{figure}[htbp]
 \centering\figfile{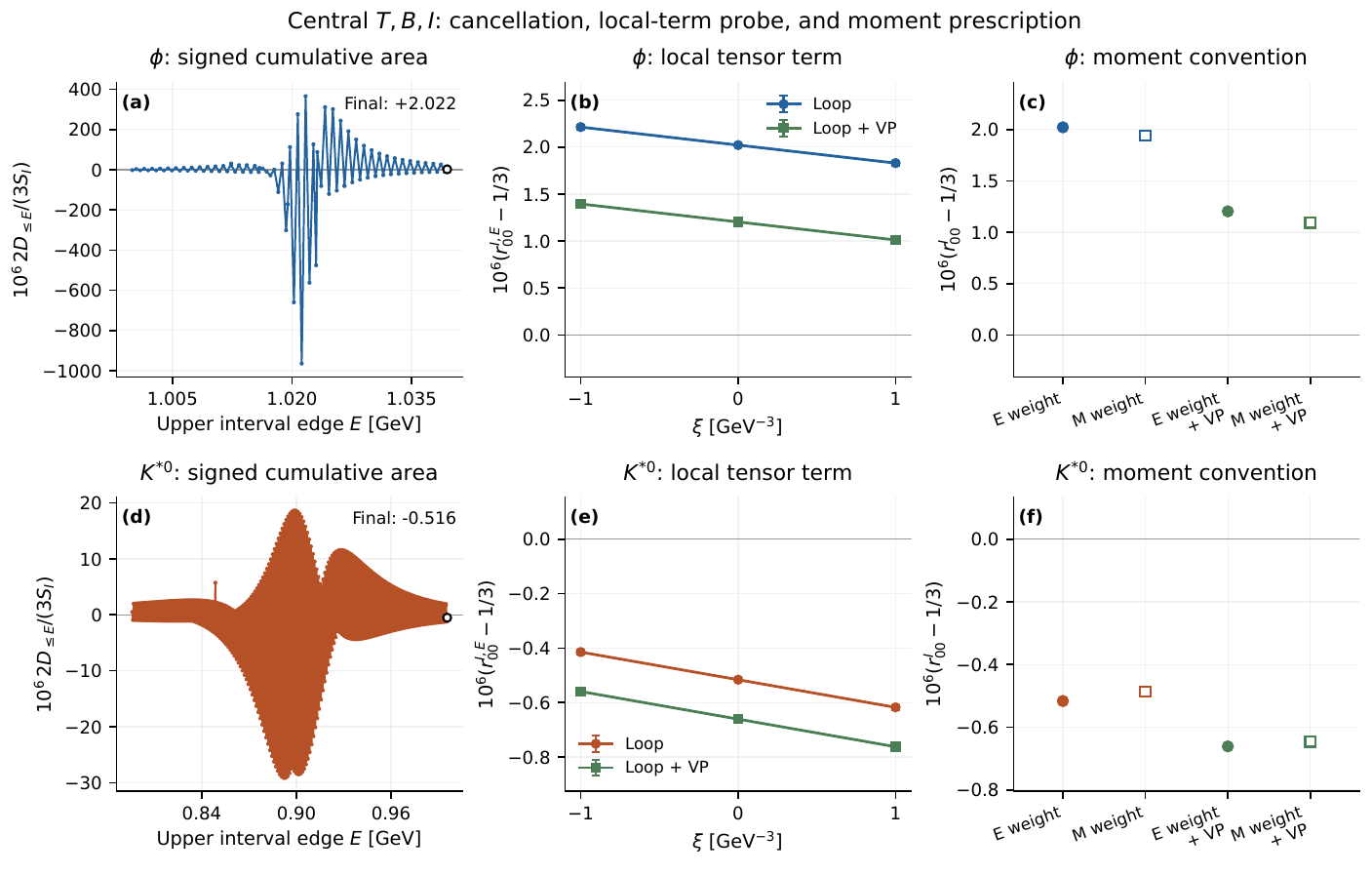}{1}
 \caption{Accumulated signed tensor area and sensitivity of the moments
at the central parameters. Left: cumulative integrals at converged
interval boundaries, normalized by the scalar integral over the full
window. The intermediate values are therefore not ratios for smaller
windows. Middle: response to the specified local term
$\Delta\Sigma_0=2M\xi b^2$, with and without $VP$ mixing. The chosen
$\xi$ range is a conditional probe and does not define a calibrated
uncertainty band. Right: comparison of energy-weighted and
fixed-reference-mass-weighted moments at $\xi=0$, with the $VP$ choice
shown separately. Numerical error bars have the restricted meaning
specified in Fig.~\ref{fig:scans}.}
 \label{fig:sensitivity}
\end{figure}

\subsection{Smooth acceptance and weak-field behavior}
\label{sec:smooth}
To examine the role of sharp boundaries separately from the accepted
energy range, we replace the indicator of $I$ in
Eq.~\eqref{eq:momentE} with a compact $C^\infty$ weight $w(E)$.
We denote the outer half-support by $H_{\rm out}$ and the width of
the edge transition by $\Delta$. Defining $d=H_{\rm out}-|E-M|$,
we use
\begin{equation}
 w(E)=\begin{cases}
 0,&d\leq0,\\
 S(d/\Delta),&0<d<\Delta,\\
 1,&d\geq\Delta,
 \end{cases}\qquad
 S(t)=\frac{e^{-1/t}}{e^{-1/t}+e^{-1/(1-t)}}.
 \label{eq:smoothweight}
\end{equation}
The hard window is recovered at $\Delta=0$. In the fixed-support
family, we keep $H_{\rm out}=20\,\MeV$ for $\phi$ and
$100\,\MeV$ for $K^{*0}$, so that the integrated acceptance decreases
as $\int w\dd E=2H_{\rm out}-\Delta$. To control for this reduction,
we also consider a fixed-area family with $H_{\rm out}=H_0+\Delta/2$,
where $H_0=10\,\MeV$ and $50\,\MeV$ for the two species,
respectively. The symmetry $S(t)+S(1-t)=1$ then gives
$\int w\dd E=2H_0$, independently of $\Delta$.
We use six transition widths for each family. The weights coincide at
the largest transition, which provides an internal consistency check.

We include each weight directly in the adaptive integral of the complete
stable-daughter spectrum and resolve the narrow peaks at the same time.
This procedure uses neither midpoint reweighting nor spectral smoothing.
The resulting quantities are field spectral moments for specified
acceptances; the weights do not describe a detector or a physical
source. The daughters remain stable in these scans. We examine a
specified endpoint broadening for a smooth weight separately in
Sec.~\ref{sec:widthbridge}.

For $\phi$ at the central field, smoothing inward over the full support
($H_{\rm out}=\Delta=20\,\MeV$) changes the loop result from
$+2.0216$ to $-0.59189$ ppm. The fixed-area result starts at
$-37.5878$ ppm for $\Delta=0$, changes to $+2.63$ ppm at
$\Delta=2\,\MeV$, and reaches the same $-0.59189$ ppm at the
common endpoint. Since the fixed-area family also exhibits this
variation, the smoothing dependence cannot be attributed solely to a
loss of accepted area. Nor does smoothing select a universal sign for
arbitrary weight shapes.

\begin{table}[htbp]\centering\small
\caption{Results for the smoothest $\phi$ acceptance considered,
$H_{\rm out}=\Delta=20\,\MeV$, at $T=150\,\MeV$. Deviations and
numerical indicators are given in ppm. The indicators combine
quadrature estimates with omitted-strip bounds and exclude model
errors.}
\label{tab:smooth}
\begin{tabular}{rrrrr}\toprule
$b/(10^{-4}\,\GeV^2)$ & Loop & Loop+$VP$ & Numerical indicator & Loop $\delta r/b^2$ [$\GeV^{-4}$]\\\midrule
1.25 & -0.14810 & -0.55866 & 0.0175 & -9.4785 \\
2.5 & -0.59189 & -2.23415 & 0.0101 & -9.4703 \\
5 & -2.43147 & -9.00072 & 0.0064 & -9.7259 \\
10 & -6.25515 & -32.53549 & 0.0045 & -6.2551 \\
\bottomrule\end{tabular}\end{table}

The field dependence for the most smoothed $\phi$ weight is given in
Table~\ref{tab:smooth}. At the two weakest fields, the loop model gives
$(r_{00}-1/3)/b^2=-9.4785,-9.4703\,\GeV^{-4}$.
The corresponding quadrature-plus-strip indicators are about $1.12$
and $0.16$ in these units, exceeding the difference between the two
central coefficients. The sampled results are therefore consistent
with a quadratic response for this fixed smooth weight. At
$b=10^{-3}\,\GeV^2$, however, the coefficient is
$-6.2551\,\GeV^{-4}$, and a single coefficient no longer describes
the full scan. Agreement at the two weakest points does not establish
an asymptotic expansion as $b\to0$. Moreover, including the specified
$VP$ term changes these two coefficients to
$-35.7545,-35.7465\,\GeV^{-4}$. The dependence on analytic matching
terms therefore remains after smoothing the energy projection.

For $K^{*0}$ at the central field, the fixed-support loop result changes
from $-0.516368$ ppm for a hard window to $+0.055880$ ppm at
$\Delta=100\,\MeV$. In the fixed-area family, the transition widths
$\Delta=(0,5,10,25,50,100)\,\MeV$ give
$(6.01932,\allowbreak0.15913,\allowbreak0.05724,\allowbreak0.05682,\allowbreak0.05663,\allowbreak0.05588)$ ppm.
The smoothed results vary much less than the hard-edge member. This
relative stability does not remove the sensitivity to known matching
contributions: the fully smoothed loop+$VP$ result is $-0.14099$ ppm,
which has the opposite sign. The numerical indicator for the fully
smoothed loop result is $0.0054$ ppm.

To check the smooth integrals, we increase the number of root probes
for the central $\phi$ calculation from 22 to 32 and tighten the
quadrature tolerances. The 24 ratios, comprising 12 weights with and
without $VP$, change by at most $7.0\times10^{-15}$.
The same refinement for the central $K^{*0}$ calculation changes them
by at most $8.9\times10^{-16}$, and all 72 regression, common-weight
and refinement checks are satisfied. We retain the conservative
endpoint-strip bounds in Table~\ref{tab:smooth}; thermal-cache
residuals and uncertainties from the physical truncation are separate.
These convergence results do not imply that acceptance smoothing can
replace daughter dressing or source matching.

\begin{figure}[htbp]
 \centering\figfile{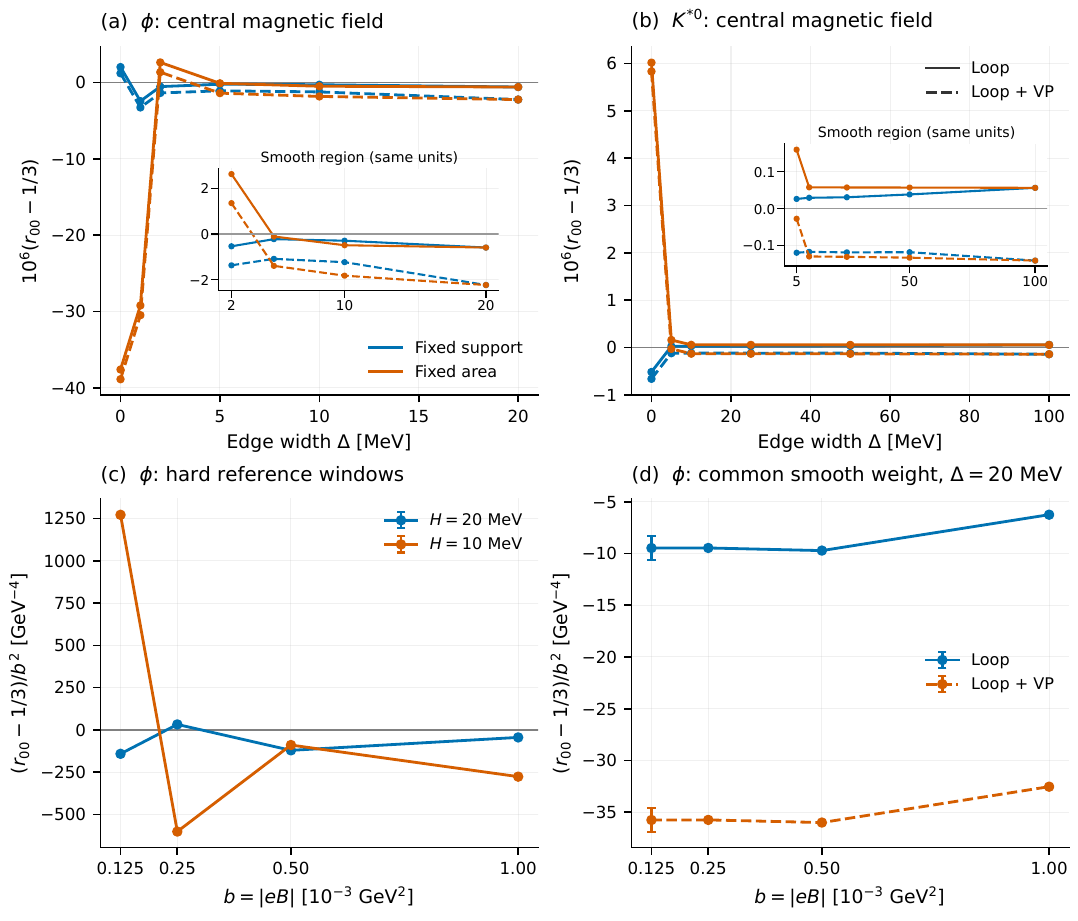}{1}
 \caption{Spectral moments with smooth weights and unchanged stable
daughters. Panels (a,b) compare the compact fixed-support and fixed-area
families at $b=2.5\times10^{-4}$ GeV$^2$, with and without $VP$
mixing. The insets show the region of smooth weights in greater detail.
The fixed-support half-windows are 20 MeV for $\phi$ and 100 MeV for
$K^{*0}$. The fixed-area families have half the corresponding integrated
acceptances and begin with narrower hard windows. Panel (c) gives the
hard-window $\phi$ ratio $(r_{00}-1/3)/b^2$ at four fields.
Panel (d) shows the common weight at $\Delta=20$ MeV, including the
$VP$ comparison. Error bars combine quadrature estimates and bounds on
omitted strips, and the connecting lines are not fits. Smoothing reduces
sensitivity to sharp boundaries, but does not give acceptance-independent
coefficients or determine the $K^{*0}$ sign in the presence of $VP$
mixing.}
 \label{fig:smooth}
\end{figure}

\subsection{Threshold redistribution and the smooth spin moment}
\label{sec:widthbridge}
We now study how the local width effect enters the accepted moment at
the central $\phi$ point, using the fixed weight
$H_{\rm out}=\Delta=20\,\MeV$. We modify the three endpoints
$\mathcal J=\{13,17,21\}$ to determine whether the loss of their
narrow structures produces a comparable fractional change in the
smooth spin ratio. This is a specified sensitivity calculation and
does not constitute a self-consistent dressing of all kaon lines.

We choose the same endpoints whose peak areas were tested with the
complete stable inverse in Sec.~\ref{sec:width}. Their detunings are
$x_*=8.07$--$50.44$ eV, with $Q/D=0.0131$--$0.0553$ and
$D=0.0284$--$0.0446\,\GeV^2$. They cover a range of local peak
scales for which the area approximation has already been checked and
the regular background remains finite. The selection is a diagnostic
subset, without a claim of completeness or statistical representativeness.
At this field, the support $I$ contains 40 transverse thresholds,
$r=13,\ldots,52$. The nearest threshold to the vacuum parent mass,
$r=32$ at $E_t=1.019248586$ GeV, is left unchanged because the
near-parent-pole regime lies outside this local calculation.
Table~\ref{tab:endpointselection} lists the selected energies and
acceptance weights. For $\Delta=5$ MeV, the levels $r=17,21$ have
essentially full acceptance. This tests a different weighting from
the strong edge suppression of $r=13$ at $\Delta=20$ MeV.
All response indicators below refer to these three replacements and
place no constraint on the combined effect of dressing every threshold
or the regular terms.

\begin{table}[htbp]\centering\small
\caption{Positions and acceptance weights of the three retained
transverse endpoints. Both weights use $H_{\rm out}=20$ MeV. The
selected endpoints are the local benchmarks whose areas were validated
previously; they do not include all 40 thresholds in $I$.
For $r=17$, the 5 MeV weight is exponentially close to unity.}
\label{tab:endpointselection}
\begin{tabular}{rrrr}\toprule
$r$ & $E_t-M_\phi$ [MeV] & $w(E_t)$, $\Delta=20$ MeV
& $w(E_t)$, $\Delta=5$ MeV\\\midrule
13 & $-19.026$ & $3.439\times10^{-9}$ & $0.01998$\\
17 & $-15.036$ & $0.06304$ & $\simeq1$\\
21 & $-11.061$ & $0.39427$ & $1$\\
\bottomrule\end{tabular}\end{table}

To relate local redistribution to the accepted moment, we introduce
disjoint patches $P_j=[E_j-L_j,E_j+L_j]$ and define a signed measure,
its first moments, and an absolute second moment:
\begin{align}
 \dd\mu_j(E)&=\frac{E}{\pi}[A_h(E)-A_0(E)]\dd E,
 &f(E)&=w(E)n_B(E),\\
 \Delta Z_j&=\int_{P_j}\dd\mu_j,
 &\Delta M_{1,j}&=\int_{P_j}(E-E_j)\dd\mu_j,\\
 T_{2,j}&=\int_{P_j}(E-E_j)^2|\dd\mu_j|.
\end{align}
One may use either the local spectral difference or the difference
from the full-background intervention defined below, as long as the
same spectra enter every term. Expanding the smooth weight by Taylor's
theorem gives the exact relation
\begin{align}
 \Delta N_{T,P_j}
 &=f(E_j)\Delta Z_j+f'(E_j)\Delta M_{1,j}+R_{2,j},\\
 |R_{2,j}|&\leq\frac12\sup_{P_j}|f''|\,T_{2,j}.
 \label{eq:weightedremainder}
\end{align}
The remainder inequality is analytic, while $T_{2,j}$ is evaluated
by quadrature with a separate convergence estimate. Because the
zeroth moment is signed, its smallness does not alone limit the
weighted response. The first moment and the contribution outside
the patches must also be examined. We obtain the derivative bounds
analytically over each interval, rather than by taking the largest
value on a sampled grid. For the constant-$D,Q$ local model, we test
the Taylor relation against direct weighted integrals with
$L_j=\Delta E_{{\rm LL},j}/(32,16,8)$. The calculation with the full
background separately integrates its own spectral difference.

To include the regular background and specify the continuation outside
the patches, we define $\delta K_h(a)=K_h(a)-a^{-1/2}$ and use
\begin{align}
 F_{0,h}(z)&=F_{0,0}(z),\\
 F_{\pm,h}(z)&=F_{\pm,0}(z)+
 \sum_{j\in\mathcal J}\mathcal C_j
 \left[\delta K_h(E_j-z)+\delta K_h(E_j+z)\right].
 \label{eq:endpointintervention}
\end{align}
Here $F_{m,0}$ is the complete matched inverse for stable daughters;
the constant-$D,Q$ approximation is no longer used. The second term
in brackets supplies the negative-frequency mirror. With conjugate
rim prescriptions, it ensures $F_m(-z^*)=F_m(z)^*$ in this neutral
equal-mass channel. The added self-energy is analytic in the upper
half-plane and extends without a real-energy patch cutoff. Analyticity
of this addition alone, however, does not exclude upper-half-plane
poles of the resummed propagator.

For $|a|\gg h$, symmetry of the pair distribution leads to
\begin{equation}
 \delta K_h(a)=\frac{h^2}{4a^{5/2}}
       +O\!\left(\frac{h^4}{a^{9/2}}\right).
 \label{eq:endpointtail}
\end{equation}
The replacement therefore introduces no ultraviolet subtraction
polynomial, and we keep the original matching coefficients fixed.
Both the regular continuation and the selection of endpoints are
explicit choices of this calculation. Although the added term generally
changes the finite value and slope at the original subtraction point,
we do not perform a compensating rematch. Thus this construction does
not calculate the width-dependent regular terms, contacts or vertices
of the full hadronic loop. The daughter convolution is performed before
the parent Dyson inversion. A convolution of the parent spectrum or
$r_{00}$ would describe a different operation
\cite{CrivellinHoferichter2023}.

We can establish positivity over the accepted energy interval
analytically. The exact stable open-channel term has the form
$C_j(E)/\sqrt{E-E_j}$, with coefficient
\begin{equation}
 C_j(E)=\frac{g^2b^2j\,[1+n_B(e_a)+n_B(e_c)]}
 {\pi\sqrt{(E+E_j)[E^2-(u_j-v_j)^2]}},\qquad
 e_{a,c}=E/2\mp b/E.
\end{equation}
This coefficient decreases above threshold. Provided that
$0\leq-C'_j(E)\leq C_j(E_j)\ell_j$, removing the leading stable
endpoint leaves a remainder no smaller than
$-C_j(E_j)\ell_j\sqrt{E-E_j}$.
Bounding each term in $-\partial_E\log C_j$ limits the total negative
remainder to $4.320\times10^{-6}\,\GeV^2$.
The unchanged neutral vacuum cut is at least
$1.0692\times10^{-4}\,\GeV^2$ over $I$, and the broadened
positive-frequency kernels contribute nonnegative absorption. For the
tested widths, the mirror terms are real on $I$ when $h\leq1$ keV.
Combining these contributions gives
\begin{equation}
 \Imm F_{\pm,h}(E)\geq1.0260\times10^{-4}\,\GeV^2>0,
 \qquad E\in I.
 \label{eq:bridgepositive}
\end{equation}
This lower bound also controls the weight omitted in numerical endpoint
strips. It establishes positivity on the rim within the chosen interval,
without providing a global count of propagator poles.

We obtain the modified spectrum from $A_{m,h}=-2\Imm F_{m,h}^{-1}$
and integrate its difference from the stable spectrum over the full
compact support. Since the three replacements enter the same inverse,
their interference is retained. Introducing $S=\sum_mN_{m,0}$,
$r=N_{0,0}/S$, and $d=N_{+,h}-N_{+,0}=N_{-,h}-N_{-,0}$,
we find the exact change in the ratio, with the longitudinal inverse
held fixed,
\begin{equation}
 \Delta r_{00}[w;h]=-
 \frac{2r\,d}{S+2d},\qquad
 d=\sum_j\Delta N_{T,P_j}+\Delta N_{T,I\setminus\cup_jP_j}.
 \label{eq:bridgeratio}
\end{equation}
We evaluate this expression without expanding in $d/S$. If the error
in $d$ is bounded by $\varepsilon$ and $S+2d>2\varepsilon$, the
associated ratio error is bounded by
\begin{equation}
 \frac{2rS\varepsilon}{(S+2d)(S+2d-2\varepsilon)}.
 \label{eq:bridgeerror}
\end{equation}
The stable baseline is common to both calculations. We propagate its
uncertainty through their difference, instead of treating the nearly
equal ratios as independent measurements.

We first apply the Taylor relation to 36 local cases, comprising three
endpoints, four widths and three patch sizes, and perform nine additional
checks with tighter quadrature. In every case the Taylor residual is
smaller than the analytic remainder estimate. At $h=1$ keV and
$L_j=\Delta E_{{\rm LL},j}/8$, for example, the three patches give
the following contributions per transverse mode:
\begin{align}
 \sum_j f(E_j)\Delta Z_j&=-3.45435\times10^{-13},\\
 \sum_j f'(E_j)\Delta M_{1,j}&=-2.99515\times10^{-14},\\
 \Delta N_{T,P}^{\rm local}&=-3.75155\times10^{-13},
 &\sum_j|R_{2,j}|&\leq1.069\times10^{-15}.
\end{align}
The first moment supplies about 8\% of this small local response and
cannot be omitted on the basis of area redistribution alone. Direct
integration yields $\Delta r^{\rm local}=9.3912\times10^{-11}$,
with Taylor remainder bound $2.675\times10^{-13}$ and quadrature
estimate $6.7\times10^{-19}$. These results apply to the local patches;
the full-window intervention is evaluated separately.

For the full-background calculation, we choose
$L_j=\Delta E_{{\rm LL},j}/16$ and include all energies in $I$
outside the patches. We resolve both unchanged and modified roots,
as well as the edges of the box support. Comparing orders 24 and 48
tests quadrature convergence. An order-64 calculation with the endpoint
guard increased from $2\times10^{-13}$ to $4\times10^{-13}$ GeV
then tests the combined quadrature and guard dependence. We evaluate
the algebraic propagator difference directly to avoid subtracting two
nearly equal full moments. Equation~\eqref{eq:bridgepositive} bounds
the contribution of excluded strips. Outside the modified supports,
we obtain the stronger estimate
\begin{equation}
 |\delta K_h(x)|\leq\frac{h^2}{4(|x|-2h)^{5/2}},\qquad |x|>2h,
\end{equation}
from the vanishing first moment of the symmetric pair distribution
and a bound on the second derivative. Together with the
propagator-difference identity, this bounds the omitted difference
itself, without separately bounding the two complete spectra at every
threshold.

\begin{table}[htbp]\centering\small
\caption{Full-window response budget for the intervention on three
endpoints, with $H_{\rm out}=\Delta=20$ MeV. The stable-daughter signal
is $r_{00}-1/3=-0.591891$ ppm. The quantity $\mathcal U_r$ combines an
analytic omitted-strip bound with numerical convergence indicators;
it is not an interval-certified bound on the entire quadrature.
The final column gives the sum of the numerical indicators included
in $\mathcal U_r$.}
\label{tab:widthbridge}
\begin{tabular}{rrrr}\toprule
$h$ [eV] & $\mathcal U_r$ [ppm] & $\mathcal U_r/|r-1/3|$ [\%]
& $\varepsilon_{\rm num}$ [ppm]\\\midrule
10 & 0.0002010 & 0.03396 & 1.98e-08\\
100 & 0.0002020 & 0.03413 & 1.79e-07\\
1000 & 0.0002013 & 0.03401 & 1.31e-07\\
\bottomrule\end{tabular}\end{table}

To characterize the size of the calculated response conservatively,
we use the indicator
\begin{equation}
 \mathcal U_r=|\Delta r_{\rm cut}|+B_{\rm strip}
       +\varepsilon_{\rm order}+\varepsilon_{\rm guard}
       +\varepsilon_{\rm baseline},
 \label{eq:bridgebudget}
\end{equation}
where $\Delta r_{\rm cut}$ is the result with numerical endpoint
strips excluded, and $B_{\rm strip}$ bounds their ratio contribution
analytically. The remaining terms are indicators of numerical
convergence and error propagation; they are not rigorous interval
quadrature remainders. For the two smooth weights, the order and guard
indicators are at most $1.69\times10^{-13}$ and $2.11\times10^{-13}$,
respectively, in the dimensionless ratio. Both are much smaller than
the strip bound, which supplies more than 99\% of $\mathcal U_r$
in all six cases. The nearly constant dependence on $h$ consequently
comes from the conservative strip allowance and does not imply a
width-independent physical response. The data include an estimate
with the strips restored, but the sign of that much smaller net
response remains unresolved.

The integration outside the patches is essential to this conclusion.
For $h=1$ keV and $\Delta=20$ MeV, the complement contributes
$+1.04483\times10^{-12}$ to $\Delta N_T$, almost cancelling the
patch sum of approximately $-1.04\times10^{-12}$. Omitting this
term would therefore give the wrong magnitude of the net response.
The full-background calculation includes the whole window explicitly;
varying the local patch size cannot replace its complement.

At $h=100$ eV, all three local real-zero merger values in
Table~\ref{tab:widthscales} have already been exceeded, while
$2h/\Delta E_{\rm LL}\simeq2\times10^{-4}$ remains small.
Nevertheless, Table~\ref{tab:widthbridge} gives
$\mathcal U_r<2.03\times10^{-10}$, or 0.0343\% of the stable
smooth signal, over the sampled range of 10--1000 eV. For the second
weight, with $\Delta=5$ MeV, we find $\mathcal U_r<1.013\times10^{-9}$,
below 0.453\% of its stable signal. These fixed-field comparisons show
that the loss of the selected narrow real zeros need not produce a
comparable change in the smooth moment. They are not a weak-field
scaling fit, a bound on dressing all thresholds and width-dependent
regular terms, or a determination of a source-independent spin
observable.

\begin{figure}[htbp]
 \centering\figfile{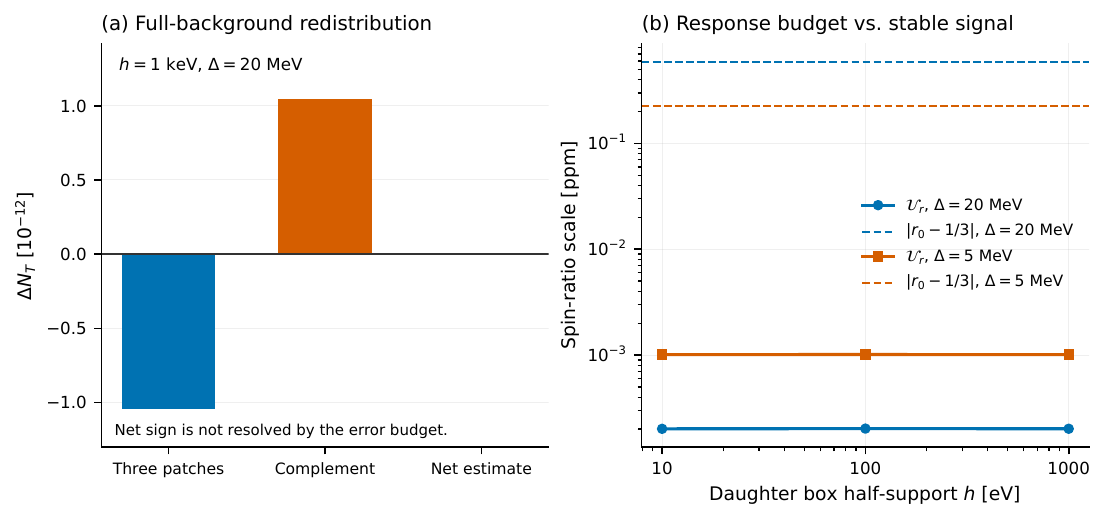}{1}
 \caption{Effect of endpoint redistribution on the smooth accepted spin
moment at the central $\phi$ point. (a) Contributions from three patches
of half-size $\Delta E_{\rm LL}/16$ and their complement in the full
background at $h=1$ keV and $\Delta=20$ MeV. The patch and net
estimates include the estimated restoration of the tiny numerical
endpoint strips. The conservative budget does not resolve the sign of
their sum. (b) Response indicator $\mathcal U_r$ from
Eq.~\eqref{eq:bridgebudget}, in ppm, compared with the magnitude of
each stable smooth signal. The analytic strip bound dominates the
combined indicator and accounts for its nearly flat width dependence;
this trend is not a measured physical width law. Both panels show the
three-endpoint intervention of Eq.~\eqref{eq:endpointintervention},
rather than a complete loop with dressed daughters.}
 \label{fig:widthbridge}
\end{figure}

\subsection{Role of the real self-energy terms}
\label{sec:ablation}
To examine the contributions of the real self-energy, we compare the
complete central model with three deliberately incomplete inverses.
We separately remove the thermal scattering terms of both charged and
neutral channels, their thermal connection contact, and only the common
magnetic matching polynomial
$\Delta\Sigma_L(s_*)+(s-s_*)\Delta\Sigma_L'(s_*)$.
The tensor and circular magnetic matching terms, zero-field subtractions
and remaining inputs are kept fixed. These removals isolate the role
of individual terms; they neither define consistent alternative
effective actions nor provide a physical error band.

Because the removed contributions are real inside the selected resonance
windows, all four inverses satisfy the same absorptive and optical
checks there. Their real zeros and spectral moments can nevertheless
differ, and the scattering discontinuity outside the window need not
be preserved. We resolve each variant's roots and widths independently
before joint adaptive integration.

\begin{table}[htbp]\centering\small
\caption{Effect of selectively omitting real-part contributions in the
central hard windows. These deliberately incomplete calculations give
$r_{00}-1/3$ in ppm, together with indicators that combine quadrature
estimates and omitted-strip bounds. The absorptive parts are identical
inside the windows. The different rows do not define a physical
uncertainty band.}
\label{tab:ablation}
\begin{tabular}{lrr}\toprule
Inverse & $\phi$ & $K^{*0}$\\\midrule
Complete & $2.021555 \pm 0.02407$ & $-0.516369 \pm 0.00933$\\
Without thermal scattering & $2.020946 \pm 0.02407$ & $-0.519323 \pm 0.00930$\\
Without thermal contact & $0.928908 \pm 0.02279$ & $-0.270212 \pm 0.00836$\\
Without common magnetic matching & $2.021555 \pm 0.02407$ & $-0.516369 \pm 0.00933$\\
\bottomrule\end{tabular}\end{table}

For $\phi$, removing the common thermal contact changes $r_{00}-1/3$
from $2.02156$ to $0.92891$ ppm. The difference, $-1.09265$ ppm,
exceeds the conservative summed numerical indicator of $0.04686$ ppm.
Although the contact $\Sigma_{\rm ct}^T=0.00877767\,\GeV^2$ is
spin independent, it enters different spin denominators and therefore
does not cancel in the normalized ratio. The same removal for $K^{*0}$
changes the deviation from $-0.51637$ to $-0.27021$ ppm, giving a
difference of $+0.24616$ ppm against a summed indicator of $0.01769$
ppm. For both species, the changes caused by the other two removals
are smaller than their respective summed numerical indicators and
remain unresolved.

The latter conclusion is specific to the tensor ratio. For the circular
difference $P_z=(N_+-N_-)/(N_0+N_++N_-)$, removing thermal scattering
changes the central $K^{*0}$ value from $2.88239\times10^{-5}$ to
$1.90333\times10^{-5}$, a reduction of $34\%$. The difference
$-9.79058\times10^{-6}$ is larger than the summed numerical indicator
$1.86\times10^{-8}$. A term with little effect on $r_{00}$ can thus
remain relevant to another spin combination. These comparisons show
why agreement in absorption over the measurement window does not
determine the spectral ratio. They make no claim of an error in a
published calculation based on a different action.

\section{Discussion}
\label{sec:discussion}
The results obtained above describe a stable-daughter, one-loop
calculation in the specified hadronic model. The narrow peaks in
Fig.~\ref{fig:threshold} characterize this limit. Their response to
positive daughter spectral input is studied locally in
Sec.~\ref{sec:width}, and the three endpoint replacements in
Sec.~\ref{sec:widthbridge} determine the corresponding full-window
response indicator. A self-consistent thermal kaon or pion width, the
magnetic three-pion sector of $\phi$, finite parent momentum, medium
expansion, and experimental reconstruction remain outside the present
calculation. Extending the result to fully dressed daughters requires
consistent regular terms, matching conditions, and vertices together
with a microscopic daughter self-energy. The finite-domain stability
result established for the original matched inverse would also need to
be checked for such an extension.

It is useful to discuss separately what the retarded kernel implies for
time-dependent spin evolution. The matching conditions and KMS relation
do not by themselves provide a finite-time kinetic closure. In the loop
calculation, the dispersive derivative
$c_m=\partial_s\Ree F_m$ has the central values
$(1.005729,1.246149,1.246149)$ for $\phi$ and
$(0.989223,2.138073,2.508549)$ for $\Kst$.
This quantity is defined on the real energy axis and is not, in general,
a pole residue. In particular, the transverse $K^*$ values show that
setting the drift coefficient to unity would introduce an additional
approximation.

The role of the dispersive terms can be seen more generally in the
Botermans--Malfliet form of the transport equation. With $R=\Ree F$,
$\Gamma=2Q$ and $A=\Gamma/(R^2+\Gamma^2/4)$, the time-derivative
coefficient for a homogeneous system in a fixed bath contains
\begin{equation}
 A\partial_E R+\Gamma\partial_E\Ree D^R
 =\frac{A^2}{2}\left(\Gamma\partial_E R-R\partial_E\Gamma\right).
 \label{eq:bm}
\end{equation}
Thus, the width derivative and backflow enter in addition to the
derivative of the real inverse. They cannot be determined from $c_m$
alone. The original KB equation and its BM rearrangement must be applied
with their common gradient and constraint assumptions \cite{IKV2003}.
A consistent transport calculation therefore requires more than replacing
a Markov rate by $\nu/c_m$. We leave this time-dependent problem separate
from the static spectral moments considered in this paper.

\section{Summary and conclusions}
\label{sec:conclusions}
In this paper, we have studied magnetic thresholds and spin-resolved
spectral moments in a specified hadronic $\phi/\Kst$ model. We have
constructed the real-axis propagators using vacuum magnetic matching,
the connection contact, and the real part of the thermal scattering
contribution, together with the absorptive pair cuts. The calculation
reproduces published results in common limits and resolves the areas of
closed-threshold peaks that are needed for reliable spin-ratio integrals.

We have also investigated the sensitivity of these peaks to finite
daughter spectral spreads. A positive two-daughter convolution modifies
the local spectrum on the detuning scale
$x_*=(\mathcal C_t/D)^2$. The merging of zeros of the real part of the inverse propagator, the
suppression of a spectral peak, and the loss of integrated weight
describe different effects. For compact daughter spectra, the fixed-patch
integrals show substantial transfer across the threshold with little net
weight change. This conclusion is subject to the shape-dependent damping
hierarchy and the stated criteria for omitted interactions; the spectral
input does not determine a microscopic collision width.

To connect the redistribution with an accepted spin moment, we have
separated the signed area, the first energy moment, and the bounded
Taylor remainder in a weighted-moment identity. We have then evaluated
three explicit endpoint replacements in the full stable background,
including the contribution from the patch complement. This calculation
determines the response to the specified replacements directly, without
estimating the integrated change from the peak height.

Our stable-daughter results show that the energy acceptance also affects
the spin spectral moments. Comparisons at fixed smooth acceptance area
demonstrate that this dependence is not simply due to a change in the
accepted bandwidth. For a fixed smooth $\phi$ weight, the negative
response at the weakest sampled fields is compatible with quadratic
scaling. Its coefficient, however, changes substantially when the
specified $VP$ matching is included. The results therefore establish
the threshold scales and projection dependence within this model.
They do not establish an acceptance-independent sign for the model
response or determine a reconstructed experimental $\rho_{00}$. A fully daughter-dressed
calculation, together with a matched physical production or decay
operator, is needed for that phenomenological application.

\begin{acknowledgments}
The work of L. Y. is supported by the
NSFC under Grant No. 11605072 and the Seeds Funding of Jilin University.
The work of T. S. is supported by the Anhui University of Science and Technology under Grant  Grant No. 2024yjrc164.
The work of X. W. is supported by the National Natural Science
Foundation of China under Grants No. 12675170 and No. 12635010, and by the Anhui University of Science and Technology under Grant No. YJ20240001.
\end{acknowledgments}

\bibliography{references_candidate}
\end{document}